\documentclass[prb,article,showpacs]{revtex4}
\usepackage{amsmath}
\usepackage{graphicx}
\usepackage{epsfig}
\usepackage{amsfonts}
\usepackage{amssymb}
\usepackage{dcolumn}
\usepackage{bm}

\usepackage{dcolumn}

\newcommand{\h}{\mathcal{H}}

\begin{document}

\title{A HF approach to La$_2$CuO$_4$ properties:\\ hints
for matching the Mott and Slater pictures}

\author{Alejandro Cabo-Bizet$^*$ and  Alejandro Cabo-Montes de Oca$^{**}$}

 \affiliation{ $^*$Departamento de F\'{\i}sica,  Centro
de Aplicaciones Tecnol\'{o}gicas y Desarrollo Nuclear (CEADEN),
Calle 30, esq. a 5ta, La Habana, Cuba.}

\affiliation{$^{**}$Grupo de F\'isica Te\'orica, Instituto de
Cibern\'etica Matem\'atematica y F\'{i}sica (ICIMAF),\\ Calle E, No.
309, entre 13 y 15, Vedado, La Habana, Cuba.}


\date{September 28, 2008}

\begin{abstract}
It is argued that a  Hartree-Fock (HF) solution for  Coulomb
interacting electrons in a simple model of the Cu-O planes in
La$_2$CuO$_4 $, is able to predict some of the  most interesting
properties of this material, such as its insulator character and the
antiferromagnetic order. Moreover, the natural appearance of
pseudogaps in their paramagnetic and superconductor phases are also
suggested by the discussion. These results follow after the
elimination of some symmetry restrictions which are usually imposed
on the single particle HF orbitals. One of them is the
simplification of their spinor dependence  to be of the so called
$\alpha$ or $\beta$ types. This constraint, seems to strongly reduce
the largest space of orbitals corresponding to the rotational
invariant HF formulation originally introduced by Dirac. We also
remove, the demand on the HF  orbitals of having a Bloch structure
in the Bravais lattice of the crystal. This procedure allows for the
consideration of HF solutions having the same symmetry than the
known (symmetry breaking) antiferromagnetic structure of the
material. It turns out that the most stable HF solution of the
problem corresponds to an antiferromagnetic and insulating state
(IAF), which magnetization rests on the experimentally determined
direction. Encouragingly, the evaluated magnetic moment per cell is
0.67 $\mu_B$, a result that satisfactorily reproduces the
experimentally measured value: 0.68 $\mu_B$.  Another HF solution
 having a slightly higher energy arises and corresponds to a paramagnetic state
 showing a pseudogap (PPG). It follows after only imposing the Bloch stucture
 on the single particle states. Finally a third paramagnetic but metallic solution
 (no gap) is also obtained by including both of the mentioned restrictions.
 The interesting result follows that this state only  differs from the
PPG pseudogap state in the form of excited empty orbitals. That is,
the occupied single particle states and the HF energy at T = 0 for
both solutions, are identical. In general, the discussion helps to
clarify the role of the antiferromagnetic correlations in the
structure of the physics of the HTSC materials. In addition, these
initial results, indicate a promising way for start conciliating the
Mott and Slater pictures in the physics of the strongly correlated
electron systems.
\end{abstract}


\maketitle

\section{Introduction}

\label{sec1:1}

The Hubbard type of models in the theory of strongly correlated
electron systems are notably successful
\cite{mott,slater1,peierls,hubbard,anderson,anderson1,kohn,rice,gutzwiller,gutzwiller1,
imada,dagoto,yanase,vonharlingen,damascelli,bmuller,Burns,almasan,mott1,fradkin}.
In particular, it is remarkable the way they reproduce the
properties of Mott insulators, such as metal-transitions oxides and
copper-oxygen layered HTc compounds \cite{imada}. However,  the
efforts for developing approaches having more basic foundations had
not ceased, due to the expectation that they could open the way for
obtaining more exact and specific
results\cite{terakura,gutzwiller,rice}. In this sense, methods that
are grouped into the so called Band Theory picture are also known as
first principle calculations in the literature. They are electronic
structure calculations that begin with the interactions among
electrons or atoms in vacuum. The study of the band structure they
predict, in principle  should offer a road toward the effective and
precise determination of the physical properties of  each
material\cite{matheiss,terakura}. Some of them are: the
Configurations Interaction scheme (CI); the Local Density
Approximations method (LDA)\cite{kohn1}, the Local Spin Density
Approximations procedure (LSDA) and Hartree-Fock method (HF).
However, the above mentioned potentialities of those first
principles approaches had been failing in  describing many of so
called strongly correlated electron system\cite{imada}. For example,
the LSDA, a sophisticated generalization of the LDA procedure, was
devised to describe local spin structures\cite{terakura}. However,
although the method  had offered satisfactory descriptions of the
physical results in few materials, this success had not been
universal and it also wrongly predicted the properties  of some
compounds, by example,  the here considered La$_{2}$CuO$_4$.

 The motivation of the present work arose from a primary  suspicion that
perhaps the self-consistent Hartree Fock (HF) method, could had been
underestimated in its possibilities for helping in the above
described searches\cite{dirac,slater1,slater}. In this sense, it can
be firstly remarked that it is widespread the criteria that for
obtaining behaviors such as the Mott insulator character, it becomes
necessary the presence of short range correlations among electrons
with spin quantized in different directions. By example,
paraphrasing one type of Mott's argument for specific systems:"...
two electrons with spin resting on contrary directions are forbidden
to occupy the same Wannier orbital... ". On another hand, the
orthodox HF approaches does not have in consideration the
correlations among electrons of different spins. Therefore two
electrons with opposite spins do not disturb one each other and
consequently both of them can occupy the same Wannier orbital.
Clearly, the usual HF approach, seems  not to be viable for
investigating a system in which the  Mott's argument is appropriate.
However, the physical sources of the validity of cited Mott's
statement in some systems are not completely clear. By example:
which is the physical origin of these short range correlations
assumed in it?. Even, the proper concept of correlations, roughly
described as: ''everything missing in the single-particle  HF state
for to be the real many body ground state'',  makes clear how
clouded their origins are. As the result of the study presented
here, we believe that many of the so called correlation effects, can
be effectively described even in the framework of the HF scheme,
after removing certain symmetry restrictions which obstacle the
finding of the best HF solutions. Such constraints  are usually
imposed on the space formed by the single particle orbitals, which
are employed to construct the determinant like states among which
the HF one shows minimal energy. By example, if after solving the HF
problem, it occurs that the resulting self-consistent potential
breaks the symmetry of the original crystalline lattice, it could
create a gap and thus produce a Mott kind of  insulating solution.
This effect was originally discovered by Slater in Ref.
\onlinecite{slater1}.  This symmetry breaking effect has been also
more recently underlined and deepened in Ref. \onlinecite{const1}.
However, the removal of this kind of symmetry restrictions alone had
not been able to describe the insulator properties of a large class
of materials\cite{terakura,imada}. One of the central results of the
present investigation, as it will be described just in what follows,
is the identification of another important kind of symmetry
restrictions that seemingly had been overlooked for long time. It
can be cited here  that a fully unrestricted formulation of the HF
problem was early done by Dirac in Ref. \onlinecite{dirac}.

 This work  will consider the Hartree-Fock  self-consistent problem
 as applied to a simple one band model of the La$_{2}$CuO$_4$\cite{pickett}, but following
 an unusual way. In order to leave freedom to obtain paramagnetic,
ferromagnetic and antiferromagnetic solutions in the same context,
we look for single particle orbitals   being non separable in their
spacial and spinor dependence, i.e. they will have the structure
$\phi(x,s)\neq\phi(x)\psi(s)$. In other words, in those states there
is no  an  absolute common  quantization direction for the electron
spin.  Thus, in each position the spin is quantized in a specific
direction, and the equations of motion to be used  will reflect this
fact. Note, that to proceed in this way is not other thing that to
apply the Dirac's unrestricted formulation of  the HF
procedure\cite{dirac}. We think that the restriction to $\alpha $
and $\beta$ types of  orbitals, usually employed in HF electronic
band and quantum chemistry calculations, prohibits from the start
the prediction of possible spontaneously symmetry breaking effects
\cite{szabo,slater}.  Such a particular structure, excessively
reduces the space of functions to be examined and consequently
annihilates some possibilities to obtain exotic solutions (like that
ones that are present in the strong correlation effects). We believe
that in the context of the band theory, or  more precisely, under
the HF approach, it could be possible to reproduce the main
characteristics of a wide class of {\em Mott insulator} kind of
materials. The results of the simple model investigated here, as it
will be seen, support this possibility for important compounds such
as La$_2$CuO$_4$, which under certain doping levels turn out in high
temperature superconductors\cite{pickett}. The present work heads in
that direction, with eyes in also showing  the potentialities of
implementing the above considerations in self-consistent HF
calculations, not only for describing electronic bands and the
associated ground states, but also for studying atomic and molecular
structures.

The exposition proceeds as follows. In the Section \ref{chap:2} we
describe the details of the  HF self-consistent method to be
employed in next sections. Specifically, it will be  discussed the
imposition of restrictions on the space of single particle states in
which the solutions will be searched and the possible physical
consequences they could carry on. In Section \ref{chap:3} the
effective model we are going to employ will be exposed. The symmetry
restrictions to be assumed and its corresponding tight-binding Bloch
basis will be defined. Section \ref{chap:4} is devoted to derive the
HF equations associated to interacting electrons which free
hamiltonian is given by the specially constructed effective tight
binding model. The tight binding Bloch basis associated to the model
is chosen in this section  to have the maximal symmetry given by the
group of translations leaving invariant the copper-oxygen plane in
La$_2$CuO$_4$.  Considering the previously mentioned paramagnetic
solution in a generic form, the free parameters of the effective
hamiltonian are adjusted to reproduce the
 form of the only half filled band in the Matheiss calculation
 of the band structure of La$_2$CuO$_4$ \cite{matheiss}.
 After defined the free hamiltonian of the model, the paramagnetic
 and antiferromagnetic solutions are obtained. The  Bloch basis shown in Section
\ref{chap:3} was employed.  Both solutions are also compared in this
section and the properties of all the obtained states are analyzed
in corresponding  subsections. Appendix \ref{AEleM} presents
notations, algebraic developments, constants and definitions cited
along the writing. In a final secci\'on the main conclusions of the
work are reviewed  and  various further tasks for its extension are
described.

\bigskip

\section{Rotational Invariant Hartree-Fock Method.}

\label{chap:2}

In the language of Quantum Mechanic the state of a system of N
particles is described by a function depending on each one
particle's spinor and spacial coordinates $f_n(x_1;s_1,..., x_N;s_N)
$, where $n$ represents the corresponding set of quantum numbers
\cite{fetter}. The HF approximation consists on supposing that the
above mentioned state can be expressed as a linear combination of
$N$-products of orthonormalized orbitals $\phi_{k_i}(x_i,s_i) $ with
$i=1,..., N $. Each one of these orbitals is interpreted as a single
particle state, because it defines amplitude and probability
distributions depending on a single particle coordinates. As usual,
in what follows the word coordinates will mean the spacial as well
as the spinor ones. If the  particles are fermions, the previously
mentioned linear combination is called Slater determinant
\cite{fetter,slater,szabo}.

Let
\begin{align}\label{hamiltoniano}
  \hat{\h}(x_1,...x_N)=\sum_{i}\hat{\h}_{0}(x_i)+\frac{1}{2}\sum_{j\neq i}V(x_i,x_j),
\end{align}
be the N-electrons system hamiltonian, including  kinetic plus
interaction with the environment hamiltonian $\hat{\h}_{0}$, besides
Coulomb interaction among pairs of electrons $V$. The HF equations
of motion for this system, leading the dynamic of the single
particle states in a self-consistent way, are
\begin{widetext}
\begin{eqnarray}\label{HF}
[\ \hat{\h}_0(x)+\sum_{\eta_1}\sum_{s^{\prime}}\int d^2x^{\prime}
\phi^*_{\eta_1}(x^{\prime},s^{\prime}) V(x,x^{\prime})
\phi_{\eta_1}(x^{\prime},s^{\prime})\ ]\ \phi_{\eta}(x,s)&\\
\nonumber-\sum_{\eta_1}[\ \sum_{s^{\prime}}\int d^2x^{\prime}
\phi^*_{\eta_1}(x^{\prime},s^{\prime}) V(x,x^{\prime})
\phi_{\eta}(x^{\prime},s^{\prime})\ ]
\phi_{\eta_1}(x,s)&=\varepsilon_{\eta}\ \phi_{\eta}(x,s),\ \ \ \ \ \
\end{eqnarray}
\end{widetext}
where $\eta=k_1,..., k_N $ is a label in the basis formed by  the
solutions. That is, each HF electron state is influenced by the
presence of the other ones. The self-consistent hamiltonian  has two
components, the coulomb like type of mean potential which their
fellows create: the direct potential and the contribution reflecting
the fact that two electrons cannot occupy the same state: the
exchange potential \cite{dirac}.

The HF energy of the $N$ electrons system and the interaction energy
of an electron in the $\eta$ state with the remaining ones, are
given by
\begin{widetext}
\begin{eqnarray}\label{energiat}
&&\\\nonumber
E_{HF}&=&\sum_{\eta}\langle\eta|\hat{\h}_0|\eta\rangle+\frac{1}{2}\sum_{\eta,
\eta_1}\langle\eta,\eta_1|V|\eta_1,\eta\rangle-\frac{1}{2}\sum_{\eta,\eta_1}\langle\eta,\eta_1|V|\eta,\eta_1\rangle,
\\\label{enerpar} && \\\nonumber
a_{\eta}&=&\frac{1}{2}\sum_{\eta_1}\langle\eta,\eta_1|V|\eta_1,\eta\rangle-\frac{1}{2}\sum_{\eta_1}\langle\eta,\eta_1|V|\eta,\eta_1\rangle,
\end{eqnarray}
\end{widetext}
 The  bracket notation definition
is given in Appendix \ref{AEleM} within the Subsection
\ref{sub:Brac}. It can be noted that the system of equations
(\ref{HF}) is rotational invariant because it is written without
imposing a spatially absolute direction for the spin quantization of
the single electron orbitals. This rotational invariant formulation
 of the self-consistent HF procedure was firstly introduced by Dirac in Ref.
\onlinecite{dirac}.

\subsection{$\alpha$, $\beta$ and symmetry restrictions.}\label{alphabeta}

 To solve (\ref{HF}) is a complicated task because
it is a system of  coupled integro-differential equations. The
iterative method is one of the most frequently employed for solving
this kind of systems and it is usually complemented by the
imposition of  symmetry restrictions  that simplify the space of
states to be investigated. However, the use of such constraints
could avoid the obtaining of special solutions non obeying the added
symmetry conditions. Although in some cases they could retain the
minimal energy one, the method can hide the existence of interesting
excited states and inclusive could  wrongly predict  the excitation
features in some cases, as it will be seen in what follows. A very
common symmetry restriction usually employed in band theory and
quantum chemistry calculations is to consider that single particle
solutions of (\ref{HF}) have spin quantized in a given direction in
every point of the  space \cite{matheiss,szabo}. That is
\begin{eqnarray} \label{e:alphabeta}
\phi_{\textbf{k}}(x,s)=\begin{cases}\phi^{\alpha}_{\textbf{k}}(x)\
u_{\uparrow}(s) & \text{$\alpha$ state},\\
\phi^{\beta}_{\textbf{k}}(x)\ u_{\downarrow}(s) & \text{$\beta$
state},
\end{cases}
\end{eqnarray}
where $u_{\uparrow \downarrow} $ represent the Pauli spinors with
spin up and down in a certain direction respectively. If the spacial
functions $\phi^{\alpha}_{\textbf{k}} $ and
$\phi^{\beta}_{\textbf{k}} $ are the same, the HF calculation  is
called a restricted one, if they are different, the procedure is
called unrestricted \cite{szabo}. It is interesting to investigate
the consequences of considering the possibly existence of non
separable single particle states being solutions of the HF problem.
A positive answer to this question can open a natural context for
obtaining solutions exhibiting magnetic properties and to allow
their comparison with paramagnetic ones.

Another important kind of restrictions posed on the first principles
band theory evaluations  is the a priori impositions of crystal
symmetries. To beforehand  impose a symmetry on the assumed to be
searched solution, although it could be well rooted in the studied
system, has  the risk of hiding a possible spontaneous breaking of
that invariance. This could occurs due to the reduction of the space
  orbitals in which we are searching. In that case it can turns out
that the obtained solutions will not be an absolute extremal of the
energy functional, but a conditional one, due to the fixed symmetry
constraint. For instance, let us consider  the  functional space
formed by  the allowed orbitals and the maximal subset  of orbitals
U which is invariant under a certain group of transformations
\textbf{T}. Consider also the maximal subset U$_s $ being invariant
under the group of transformations \textbf{T}$_s$ which is a
subgroup of \textbf{T}. Then, the set U$_s$ obtained from imposing
less symmetry restrictions a priori, should contain the set U.
Therefore, after to find  the  extremes of the same functional in U
and U$_s$, it could be possible to obtain different results. In this
case, the solution in U$_s$, in general, shall be the most stable of
both. However, also it could be the case  that looking for a
solution in U$_s $, an extremal function  also pertaining to U
arises as a solution. In such a situation,  such a configuration
could be found from the beginning by finding the extreme of  the
functional in U, that is, by imposing more symmetry restrictions. In
terms of the HF scheme, this could mean that  the states
corresponding to both solutions have identical occupied
single-particle states, but they  curiously might show different
sets of excited ones. Therefore, depending on the  particular
features of the material, removing  a priori imposed symmetry
restrictions on the set of allowed orbitals of the HF procedure, can
predict new properties for the excited single particle states of the
system.  Such one could be for instance  the gap appearance. This
effect can have  physical relevance after noting that at finite
temperatures the more stable state will be preferred by the system
and then the state showing a gap should be expected to be selected
at non zero temperature.

\section{Tight Binding Electron Model: "Removing Symmetries".} \label{chap:3}

In this section  the basis of the effective band model used to
describe  the dynamic of the less bounded La$_2$CuO$_4$ electrons
will be presented. The main considerations for defining the model
and the determination of its characteristic parameters are given. In
Subsection \ref{CuO} the simplified electronic model for the
copper-oxygen planes, as well the main definitions in its structure
are introduced. Subsections \ref{sec:TraslacionesSR} and \ref{TB}
are devoted to define the symmetry transformation group defining the
tight binding Bloch basis.

\subsection{Model for the  Cu-O planes.}\label{CuO}

It is known that at low temperature La$_2$CuO$_4$ is an
antiferromagnetic-insulator \cite{pickett}. However, in evident
contradiction with the experiments the Linear Augmented Plane Waves
(LAPW) technics \cite{matheiss} predicts a metal and paramagnetic
zero temperature properties for this material. Nevertheless, such
band calculation results show that the conduction electrons are
strongly coupled to the Bravais lattice centers of the copper oxygen
planes. Clearly this tight-binding behavior is determined by the
interaction of the electrons  with its surrounding effective
environment. This defines the  initial hypothesis of our model.

\begin{figure}[h]
\vspace{.1cm}
\begin{center}
\includegraphics[width=1.3in,height=1.3in]{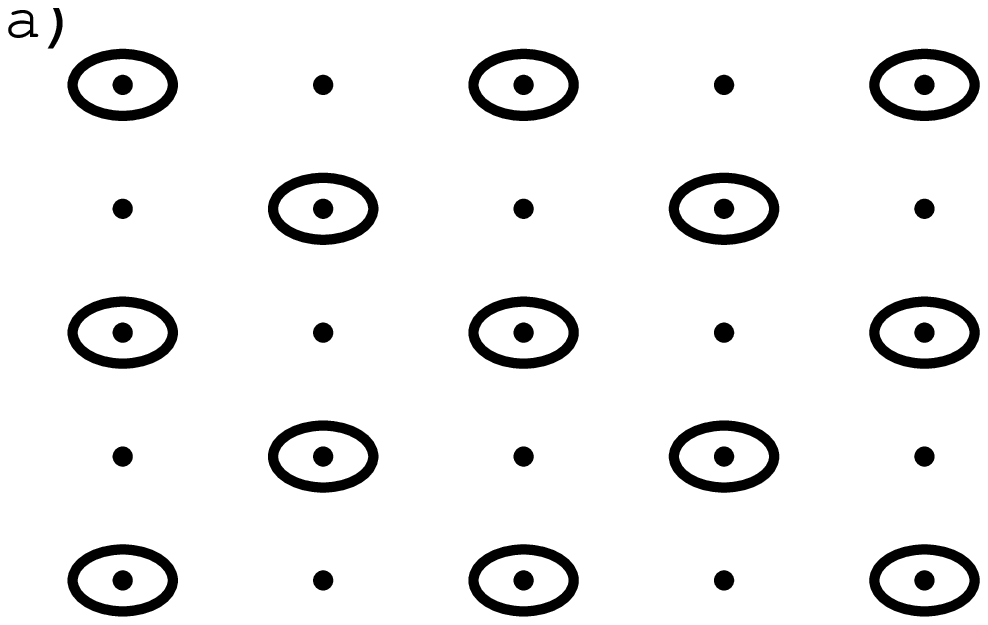}
\ \ \ \ \ \ \
\includegraphics[width=1.3in,height=1.3in]{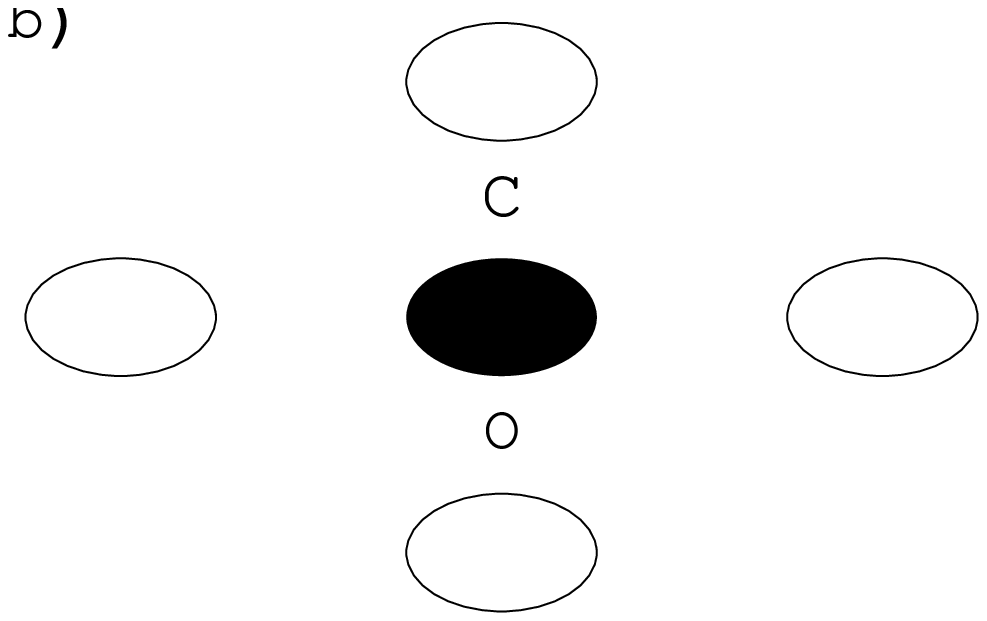}
\end{center}
\vspace{.1cm} \caption{The figures show: a) The point lattice
associated to the  Cu-O planes. For the search of the  \textbf{AFM}
properties of the conduction electron, and more generally for
removing the symmetry restrictions,   it will be helpful to separate
the lattice in the two represented sublattices; and  b) shows the
corresponding base of the Cu-O planes.}
\label{f:bravais}%
\end{figure}

The less bounded electron  in the La$_2$CuO$_4$ molecule is the
Cu$^{2+}$'s not paired one. That is, at difference from O$^{2-}$
ions,  Cu ones do not have their last shell (3d) closed. Those
copper 3d electrons fill the last band of La$_2$CuO$_4$ solid and in
what follows they shall be referred as: the electron gas. It seems
appropriate  to consider those electrons as strongly linked to
CuO$_2$ cells and moreover, given the above mentioned arguments,
with special preference for the Cu centers \cite{anderson}. Thus,
our Bravais lattice is going to be the squared net coincident with
the array of copper sites (see the figure \ref{f:bravais}).
 The presence of electrons pertaining to the various fully filled
 bands in the material plus the nuclear charges, plays a
double role in the model. Firstly, it  will act  as an effective
polarizable  environment, which screens the field created by
electron charges constituting the electron gas in the half filled
band. Consequently, we will introduce a dielectric constant
$\epsilon$ which will screen the Coulomb interaction. Secondly, as
suggested by its spacial distribution and magnitude,  the  mean
field created by the environment will be assumed to act as a
periodic potential W$_\gamma$ being responsible for tight-binding
confining of the electron to the Cu centers.

It is also primordial in the model to take into consideration the
interaction F$_b$ among the electron gas and the ''jellium''
neutralizing its charges. This background  will be modeled here as a
gaussian distribution of positive charges
\begin{eqnarray}
\rho_b(\textbf{y})=\frac{1}{\pi b^2}\exp(-\frac{\textbf{y}^2}{
b^2}),
\end{eqnarray}
surrounding each lattice point and with characteristic radius {\em b
}.

In resume, the  free hamiltonian of the model takes the form
\begin{eqnarray}\label{freehamiltonian}
\hat{\h}_0(\textbf{x})&=&\sum_{i=1}^{N}\frac{\hat{\textbf{p}}_{i}^2}{2m}+
W_\gamma(\textbf{x})+F_b(\textbf{x}),\\\nonumber
W_\gamma(\textbf{x})&=&W_\gamma(\textbf{x}+\textbf{R}),\\\nonumber
F_b(\textbf{x})&=& \frac{e^2}{4\pi\epsilon\epsilon_0}
\sum_{\textbf{R}}\int d^2y
\frac{\rho_b(\textbf{y}-\textbf{R})}{|\textbf{x}-\textbf{y}|},\ b\ll
p,
\end{eqnarray}
where $\hat{\textbf{p}}_{i}^2$ is the i-th electron's squared
momentum operator; {\em m} is the electron mass; $\epsilon_0$ is the
vacuum permitivity and
\begin{eqnarray}
\textbf{R} &=& \begin{cases}n_{x_1}p\ \hat{\textbf{e}}_{x_1}+n_{x_2}
p\ \hat{\textbf{e}}_{x_2},\\ \text{ with } n_{x_1}\text{ and }
n_{x_2}\in\mathbb{Z},\end{cases} \label{redabsoluta}
\end{eqnarray}
moves on Bravais lattice. The versors $\hat{\textbf{e}}_{x_1}$ and
$\hat{\textbf{e}}_{x_2}$ are resting on the direction defined by the
lattice's nearest neighbours, see figure \ref{f:bravais} a). It is
known that the distance between Cu nearest neighbours is p$\ \approx
\ $ 3.8 \AA \cite{pickett}. We also consider the interaction among
pairs of electrons in the form
\begin{eqnarray}\label{f:coulombefectivo}
V(\textbf{x},\textbf{y})=\frac{e^2}{4\pi\epsilon\epsilon_0}\frac{1}{|\textbf{x}-\textbf{y}|},
\end{eqnarray}
which, as remarked  above, includes a dielectric  constant
associated to the presence of the effective environment.

We are seeking here for HF solutions with orbitals having a non
separable spin and
 orbit structures. Thus, it was considered that the spin can show a
 different projection  for the different Wannier wavepackets to be superposed
 in defining those orbitals. The spin for
each of them will be either or $\alpha$ or $\beta$ type, according
they are linked either to one or  the other of the two sublattices
shown  in figure \ref{f:bravais} a). Thus, the single particle
eigenstates will be chosen to be invariant, only under the reduced
group of translations which transform each  of those sublattices on
itself.

It is important employ a procedure  leaving independently identified
the characteristics of the electron states in each one of the
sublattices. That will allow to analyze solutions with dissimilar
qualities in a same framework, that is: antiferromagnetic
\textbf{AFM}, paramagnetic \textbf{PM} and  ferromagnetic
\textbf{FM} ones. For this purpose, let us define the  points of the
two sublattices with indices r = 1, 2, as follows
\begin{eqnarray}\label{e:subred}
\textbf{R}^{(r)}&=&\sqrt{2}n_1p\ \hat{\textbf{q}}_1+\sqrt{2}n_2p\
\hat{\textbf{q}}_2+ \textbf{q}^{(r)}
\\\nonumber
&& \ \ \ \ \ \ \text{ with } n_1\text{ and } n_2\ \in
\mathbb{Z},\\\nonumber
\textbf{q}^{(r)}&=&\begin{cases}\textbf{0}, & \text{ if\ r=1},\\
p\ \hat{\textbf{e}}_{x_1}, & \text{ if r=2},\end{cases}
\end{eqnarray}
where $\hat{\textbf{q}}_1$ and $\hat{\textbf{q}}_2$ form the basis
versors on each one of them.

\subsection{Translations on the sublattices.}\label{sec:TraslacionesSR}

The solutions we are looking for will be eigenfunctions of the
operators $\hat{T}_{\textbf{R}^{(r)}}$ belonging to the reduced
discrete translation group which transforms a given sublattice on
itself:
\begin{eqnarray}\label{invariante}
\hat{T}_{\textbf{R}^{(r)}}\phi_{\textbf{k},l}&=&\exp(i\
\textbf{k}\cdot\textbf{R}^{(r)})\phi_{\textbf{k},l}.
\end{eqnarray}

\begin{figure}[h]
\begin{center}

\includegraphics[width=1.6in,height=1.6in]{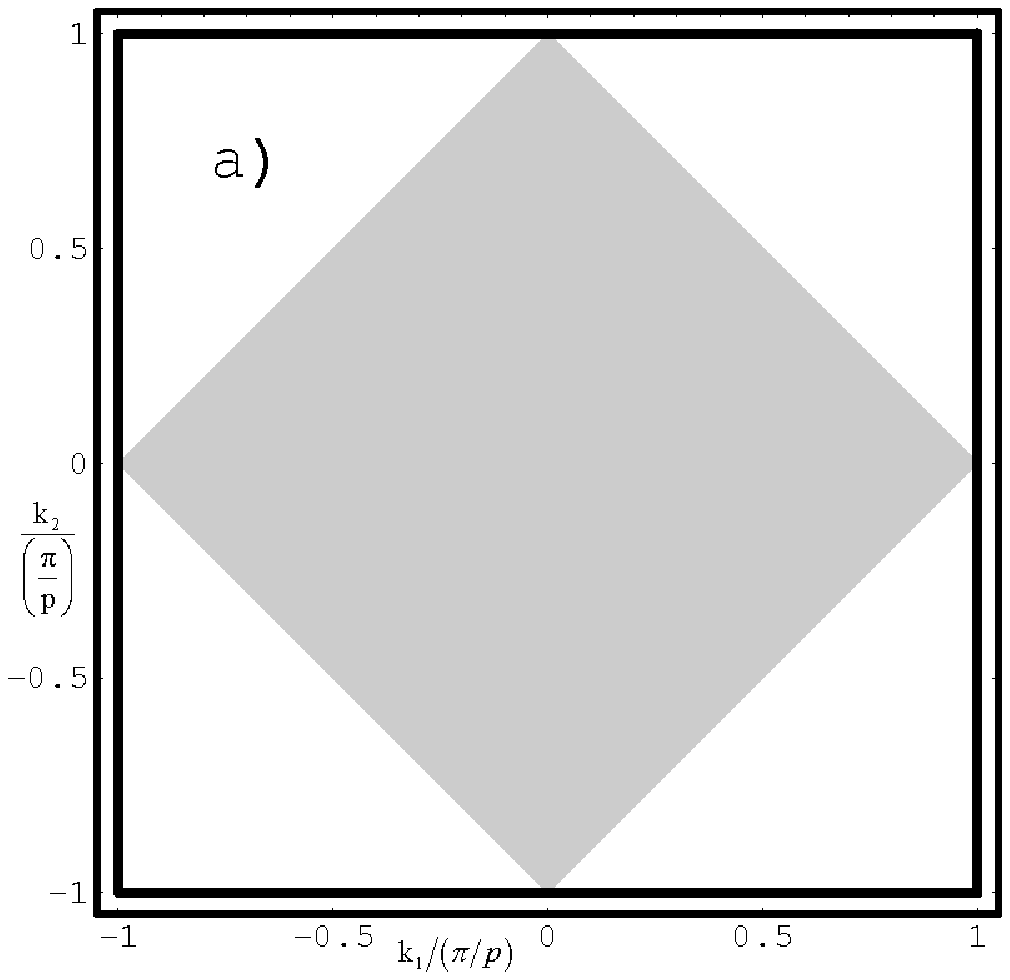}
\ \ \ \
\includegraphics[width=1.6in,height=1.6in]{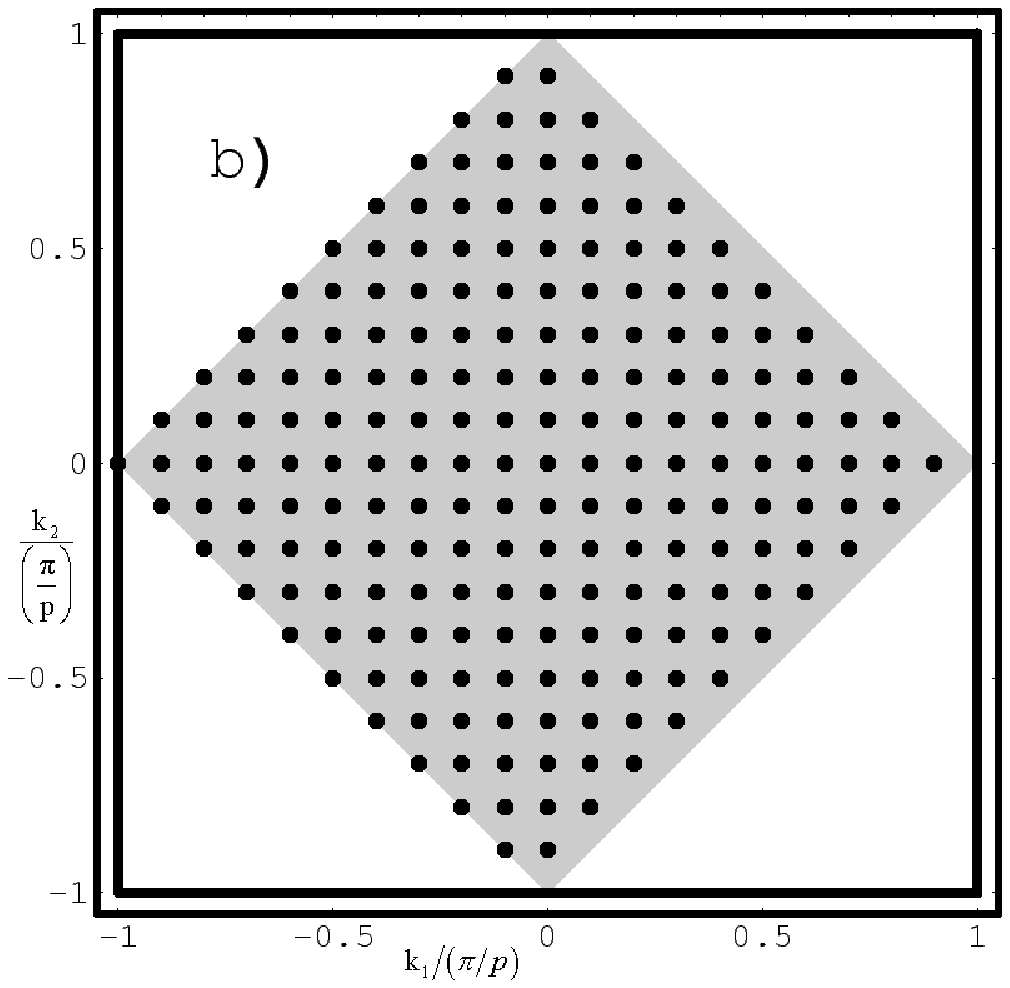}
\end{center} \caption{In the figures: a) The Brillouin zone
associated to the infinite absolute point lattice. The vectors
$\textbf{k} $ label the eigenfunctions of the group of translations
$\hat{T}_{\textbf{R}^{(r)}}$ in this infinite point lattice. The
grey square indicates the corresponding Brillouin zone (B.Z.) of the
sublattices. The length of its side is $\sqrt{2}\pi/p$.  b) The net
of points shows the discrete character of the Brillouin zone when
the absolute point lattice is finite with periodic conditions fixed
in its boundaries. The unit of the scale means $\frac{\pi}{p}$.}
\label{f:ZB}%
\end{figure}

If the Bravais lattice were infinite, the Brillouin's zone (B.Z.)
associated to $\hat{T}_{\textbf{R}^{(r)}}$ would be the shadowed one
on figure \ref{f:ZB} a). Note, that the continent square in this
figure represents the B.Z. associated to the group of translations
which leaves invariant the absolute lattice (the lattice formed by
the Cu atoms in the CuO planes). However, given the impossibility of
considering an infinite lattice for numerically solving the HF
problem, then is also not allowed to consider its associated B.Z. as
continuous. Therefore we will impose periodic boundary conditions
 on the $\phi_{\textbf{k},l}$ in the absolute lattice's
boundaries x$_1$= -L p y L p, x$_2$= -L p y L p (see figure
\ref{f:ZB} b)). This condition determines  the allowed set of
\textbf{k}
\begin{eqnarray}\label{e:ZB}
\textbf{k}&=&\begin{cases}\frac{2\pi}{Lp}\
(n_1\hat{\textbf{e}}_{x_1}+n_2\hat{\textbf{e}}_{x_2}),\\
\text{ with }n_1,\ n_2\in\mathbb{Z}\ \\ \text{and } -\frac{L}{2}\leq
n_1\pm n_2< \frac{L}{2}.\end{cases}
\end{eqnarray}

Therefore, after  recalling the discussion given in the
introduction, note that we are now demanding less crystal symmetry
on the single particle states which  we are looking for, since  a
lower number of constraints are being imposed on the space of single
particle states in which the solutions are searched.

\subsection{Tight-Binding basis}\label{TB}

The tight binding Bloch basis which we are going to use is the
following
\begin{eqnarray}\label{base de Bloch}
\\\nonumber\varphi^{(r,\sigma_z)}_{\textbf{k}}(\textbf{x},s)&=&\sqrt{\frac{2}{N}}\
u^{\sigma_z}(s)\sum_{\textbf{R}^{(r)}}\exp(i\
\textbf{k}\cdot\textbf{R}^{(r)})\
\varphi_{\textbf{R}^{(r)}}(\textbf{x}),\\\nonumber\hat{\sigma}_z
u^{\sigma_z}&=&\sigma_z\
u^{\sigma_z},\\\nonumber\varphi_{\textbf{R}^{(r)}}(\textbf{x})&=&\frac{1}{\sqrt{\pi
a^2}}\exp(-\frac{(\textbf{x}-\textbf{R}^{(r)})^2}{2\ a^2}),\ a\ll p,
\end{eqnarray}
where $N$ is the number of electrons in the  electron gas,
$\hat{\sigma}_z$ is the spin z projection operator, where z is the
orthogonal direction to the copper oxygen CuO$_2$ planes;
$\sigma_z=$ -1, 1, are the eigenvalue of the previously mentioned
operator and {\em r} = 1, 2, is the label which indicates each one
of the sublattices. As we are going to work on a half filling
condition, then $N$ coincides with the number of cells in the
crystal with fixed periodic boundary conditions $N_c$. Note, that
due to the tiny overlapping among nearest neighbors approximation,
the exact orthogonal character is only weakly lost between elements
corresponding to different sublattices and having the same spin
quantization. That occurs because nearest neighbors belong to
different arrays. However the orthogonality between different
elements corresponding to the same sublattice, as well as unity norm
for every elements, is rigourously maintained. This follows because
they are constructed as Bloch states in their corresponding
sublattices. The Wannier orbitals
$\varphi_0(\textbf{x}-\textbf{R}^{(r)})$ represent the probability
amplitude of encountering one electron in the vicinity of the site
$\textbf{R}^{(r)}$, that is on the given cell CuO$_2$.

Let us now describe  some simplifications that will be adopted in
order to solve the HF problem.  The central aim  of this work is not
to make an exact study of the problem, instead we seek for
approximate solutions qualitatively well reflecting the physical
properties of La$_2$CuO$_4$ and other compounds. Following this
principle we take for the Wannier orbitals the explicit form given
in (\ref{base de Bloch}). Physically, this means to consider the
effective potential  created by the environment on each electron of
the half filled band, is a quadratic function  having a minimum on
the  copper sites and strongly confining the electrons to it. This
last consideration is similar to the one made on the \textit{t-J}
one band model.

\section{Matrix Problem and Solutions.} \label{chap:4}
In this section  the main results of this work and their discussion
are presented. In Subsection \ref{secc:41}  the equivalent matrix
problem, resulting from projecting the HF system of equations
(\ref{HF}) on the tight binding Bloch basis (\ref{base de Bloch})
defined in the previous section is presented. Subsection \ref{SMST}
shows how  after imposing the maximal translational symmetry  and an
$\alpha$ and $\beta$ spin nature on the orbitals, the model  is able
to reproduce the dispersion relation of La$_2$CuO$_4$
metal-paramagnetic half filled band of the precise band calculations
given in Ref. \onlinecite{matheiss}. The solutions presented in
subsections \ref{solAntiFerro} and \ref{solPara}, illustrate the
consequences of releasing symmetry restrictions on the space in
which HF solutions are searched. The first of them corresponds to a
Mott's insulator-antiferromagnetic ground state. That is, the state
corresponds to an insulator even tough it has one electron per cell,
precisely in the same way as La$_2$CuO$_4$ behaves. The second
solution obtained  corresponds to a paramagnetic state showing a
pseudogap and exactly the same  HF energy and set of  occupied
single particle states,  that the previously obtained
metal-paramagnetic state\cite{matheiss}. It is important to mention
that all the band diagrams shown in this section are plotted on the
same energy scale and the zero energy point coincides with the Fermi
level of the antiferromagnetic ground state (IAF) presented in
Subsection \ref{solAntiFerro}.

\subsection{Tight-Binding representation.}\label{secc:41}

Let the searched single particle states represented in the
explicitly nonseparable form
\begin{eqnarray}\label{e:representaciontb}
\phi_{\textbf{k},\
l}(\textbf{x},s)&=&\sum_{r,\sigma_z}B_{r,\sigma_z}^{\textbf{k},\
l}\varphi^{(r,\sigma_z)}_{\textbf{k}}(\textbf{x},s),
\end{eqnarray}
where {\em l} is the additional quantum number needed for indexing
the stationary state on question, which we are going to define
precisely further ahead. After subsituting
(\ref{e:representaciontb}), (\ref{freehamiltonian}) and
(\ref{f:coulombefectivo}), in (\ref{HF}); followed by projecting the
obtained result on the basis
$\varphi^{(\textbf{t},\alpha_z)}_{\textbf{k}^\prime}$ and an
extensive algebraic work, it is possible to arrive at the following
self-consistent matrix problem for the coefficients appearing in the
expansion (\ref{e:representaciontb}):
\begin{eqnarray}\label{EcuMatricial}
[\textrm{E}^0_{\textbf{k}}+ \widetilde{\chi} \ (
\textrm{G}^{dir}_{\textbf{k}}-\textrm{G}^{ind}_{\textbf{k}}-\textrm{F}_{\textbf{k}})]\textbf{.}\textrm{B}^{\textbf{k},l}=
\widetilde{\varepsilon}_{l}(\textbf{k}) \
\textrm{I}_{\textbf{k}}\textbf{.}\textrm{B}^{\textbf{k},l},
\end{eqnarray}
where each of the quantities
\begin{eqnarray}
\textrm{B}^{\textbf{k},l}&=&\bigl\|B_{(r,\sigma_z)}^{\textbf{k},\ l}
\bigr\|,
\end{eqnarray}
represents a vector having four components given by the four
possible pairs $(r,\sigma_z)$. The constants
\begin{eqnarray}\label{e:coupling}
\nonumber\widetilde{\chi}\equiv\frac{me^2}{4\pi\hbar^2\epsilon\epsilon_0}\frac{a^2}{p},\\
\widetilde{\varepsilon}_{l}(\textbf{k})\equiv\frac{ma^2}{\hbar^2}\varepsilon_{l}(\textbf{k}),
\end{eqnarray}
are dimensionless. In them, {\em e} represents the vacuum charge of
the electron;  $\hbar$ is the reduced Planck constant; {\em a} is
the characteristic radius of the Wannier orbitals $\varphi_0$  and
{\em p} is the nearest neighbors separation. It is clear now that we
can define {\em l} = 1, 2, 3, 4, as a label indicating each of  the
 four solutions to be obtained for every value of quasi-momentum
\textbf{k}. Also, all the implicit parameters in the following
4$\times$4 matrices are dimensionless
\begin{eqnarray}
\textrm{E}^0_{\textbf{k}}&=&\bigl\|E^0_{\textbf{k},( t, \alpha_z );(
r, \sigma_z) }\bigr\|\ , \nonumber \\
\textrm{G}^{dir}_{\textbf{k}}&=&\bigl\|G^{dir}_{\textbf{k},( t,
\alpha_z ); ( r, \sigma_z) }\bigr\|\ ,\nonumber \\
\textrm{G}^{ind}_{\textbf{k}}&=&\bigl\|G^{ind}_{\textbf{k},( t,
\alpha_z ); ( r, \sigma_z) }\bigr\| \ ,\nonumber \\
\textrm{F}_{\textbf{k}}&=&\bigl\|F_{\textbf{k},( t, \alpha_z ); ( r,
\sigma_z) }\bigr\|\ ,\nonumber \\
\textrm{I}_{\textbf{k}}&=&\bigl\|I_{\textbf{k},( t, \alpha_z ); ( r,
\sigma_z) }\bigr\|\ . \label{eq18}
\end{eqnarray}
The set of quantities (\ref{eq18})  constitute the matrix
representations of the periodic potential created by the mean field
W$_\gamma$, the direct and exchange terms in (\ref{HF}), the
interaction potential with the neutralizing ''jellium'' of charges
F$_b$ defined in (\ref{freehamiltonian}) and the overlapping matrix
among nearest neighbors, respectively. Each one of the four pairs,
$(t,\alpha_z)$ and $(r,\sigma_z)$ defines a row and a column
respectively, of the matrix in question. The explicit forms of the
matrix elements are given in Appendix \ref{AEleM}.

In this new representation the normalization condition for the HF
single particle states and the HF energy of the system take the
forms
\begin{widetext}
\begin{eqnarray}
1&=&\textrm{B}^{\textbf{k},l^*}.\textrm{I}_{\textbf{k}}.\textrm{B}^{\textbf{k},l},\\
E^{HF}_{\textbf{k},l}&=&\sum_{\textbf{k},l}\Theta(\widetilde{\varepsilon}_{F}
-\widetilde{\varepsilon}_{l}(\textbf{k}))[
\widetilde{\varepsilon}_{l}(\textbf{k})
-\frac{\widetilde{\chi}}{2}\textrm{B}^{\textbf{k},
l^*}.(\textrm{G}^{dir}_{\textbf{k}}-\textrm{G}^{ind}_{\textbf{k}}).\textrm{B}^{\textbf{k},
l} ],
\end{eqnarray}
\end{widetext}
where $\Theta$ is the Heaviside function.

The system (\ref{EcuMatricial}) is non linear on the variables
$B_{r,\sigma_z}^{\textbf{k},\ l}$, which are the four components of
each vector $\textrm{B}^{\textbf{k},l}$. They can be interpreted as
a measure of the probability amplitude of finding an electron in the
state (\textbf{k}, l), in the sublattice $r$, with spin z-projection
$\sigma_z$. Thinking in numerically solving the equations by the
method of iterations, it  is convenient to pre-multiply them by
$\textrm{I}_{\textbf{k}}$ for each \textbf{k}. Note that for each
\textbf{k} four eigenvalues ({\em l} = 1, 2, 3, 4.) will be obtained
, or equivalently, four bands on the Z.B. (Eq. \ref{e:ZB}). From Eq.
(\ref{EcuMatricial}) it can be observed that in the representation
(\ref{base de Bloch}), the HF potentials and in general the total
hamiltonian of the system, resulted as  block diagonal with respect
the sets of states indexed by the same  \textbf{k}. This fact is a
consequence of the commutation of each one of them with every
element of the reduced group of discrete translations. That is, the
group of translations which leaves invariant a sublattice.

\subsection{Maximally translational symmetric solutions.}\label{SMST}

In this subsection we will  search for HF solutions  having their
orbitals on the space of Bloch functions being eigenfunction of the
maximal group of translations leaving invariant the absolute
lattice. In other words, we demand the maximum possible symmetry
under translations. It will follows that  our model is capable of
acceptably reproduce the profile of the conduction band dispersion
calculated in Ref. \onlinecite{matheiss} for the La$_2$CuO$_4$. We
will adjust the free parameters of the model in order to reproduce
in the best way  the band dispersion reported in Ref.
\onlinecite{matheiss}. The parameters are: the dielectric constant
of the effective environment $\epsilon$; the characteristic radius
of the gaussian Wannier orbitals $\widetilde{a}$; the jumping
probability between nearest sites for an electron
$\widetilde{\gamma}$ (it is fixed by the effective environment) and
 the radius in which  the gaussian orbitals
associated to the neutralizing ''jellium'' of charges decays
$\widetilde{b}$ (see Section \ref{CuO}).

 In what follows the wavy hats will mean dimensionless, see Appendix
\ref{AEleM}(\ref{e:dimensiones}). Let us define the Bloch basis for
the space of orbitals in which the solution will be searched as
\begin{eqnarray}\label{baseMatheiss}
\bar{\varphi}^{\sigma_z}_{\textbf{Q}}(\textbf{x},s)&=&\sqrt{\frac{1}{N}}\
u^{\sigma_z}(s)\sum_{\textbf{R}}\varphi^{\textbf{Q}}_{0}(\textbf{x}-\textbf{R})
\\\varphi^{\textbf{Q}}_{0}(\textbf{x})&=&\exp(i\
\textbf{Q}\cdot\textbf{R})\ \varphi_{0}(\textbf{x}),
\end{eqnarray}
where
\begin{eqnarray}\label{e:ZB}
\textbf{Q}&=&\begin{cases}\frac{2\pi}{Lp}\
(n_{x_1}\hat{\textbf{e}}_{x_1}+n_{x_2}\hat{\textbf{e}}_{x_2}),\\
\text{with }n_{x_1},\ n_{x_2}\in\mathbb{Z},\\\text{and }
-\frac{L}{2}\leq n_{x_1}\text{, } n_{x_2}< \frac{L}{2}.\end{cases}
\end{eqnarray}
are the quasimomenta of the single particle Bloch states which are
eigenfunctions of the maximal group of translations. Also is
important to define $N=L \times L$, and \textbf{R}, which are the
amount of cells in the absolute lattice and the corresponding
parametrization (\ref{redabsoluta}), respectively. The functions
$\varphi_0(\textbf{x})$ are the gaussian orbitals defined in Section
\ref{TB}. The searched HF orbitals will have the form
\begin{eqnarray}\label{estadoPMatheiss}
\bar{\phi}_{\textbf{Q},\
l}(\textbf{x},s)&=&\sum_{\sigma_z}\bar{B}_{\sigma_z}^{\textbf{Q},\
l}\bar{\varphi}^{\sigma_z}_{\textbf{Q}}(\textbf{x},s),
\end{eqnarray}
as expressed on the above mentioned basis. This time, the equivalent
matrix problem for the ''vector'' $\bar{B}^{\textbf{Q},l}$ is of
second order for each value of ($\textbf{Q}$,l). That is, its
solutions will be two component vectors. Consequently, {\em l} will
take the values 1 or 2 now. Thus, in analogy to (\ref{EcuMatricial})
the new set of equations to be solved results in the form
\begin{eqnarray}\label{EcuMatricial2}
[\bar{\textrm{E}}^0_{\textbf{Q}}+ \widetilde{\chi} \ (
\bar{\textrm{G}}^{dir}_{\textbf{Q}}-\bar{\textrm{G}}^{ind}_{\textbf{Q}}-\bar{\textrm{F}}_{\textbf{Q}})]\textbf{.}\bar{\textrm{B}}^{\textbf{Q},l}=
\widetilde{\varepsilon}_{l}(\textbf{Q}) \
\bar{\textrm{I}}_{\textbf{Q}}\textbf{.}\bar{\textrm{B}}^{\textbf{Q},l}.
\end{eqnarray}
Let $\bar{\Upsilon}$ and $\Upsilon$ anyone of the 2x2 matrices in
(\ref{EcuMatricial2}) and its 4x4 equivalent on (\ref{EcuMatricial})
respectively, the relationship between these matrix elements is the
following one
\begin{eqnarray}
\bar{\Upsilon}_{(\alpha_z,\sigma_z)}=\frac{1}{2}\sum_{t,r}\Upsilon_{(
t,\alpha_z) ; (r,\sigma_z)}.
\end{eqnarray}

The direct and exchange potentials respective matrices 2x2 and 4x4
relationship becomes  slightly more complicated. Besides also
satisfying the above mentioned relation, each vector components in
their "4x4" definitions must be  removed from the sublattice label
dependence and multiplied by $\frac{1}{\sqrt{2}}$ (those new
quantities are the vector components of the 2x2 problem). For
instance, making reference to the definitions given in Appendix
\ref{AEleM}
\begin{eqnarray}\nonumber
\bar{I}_{\textbf{Q},(\alpha_z,\sigma_z)}&=&\delta_{
\alpha_z,\sigma_z} \left[I_{00}\ +2I_{01}(\cos Q_1p+\cos Q_2p)\
\right].
\end{eqnarray}

\begin{figure}[h]
\begin{center}
\includegraphics[width=1.5in,height=1.5in]{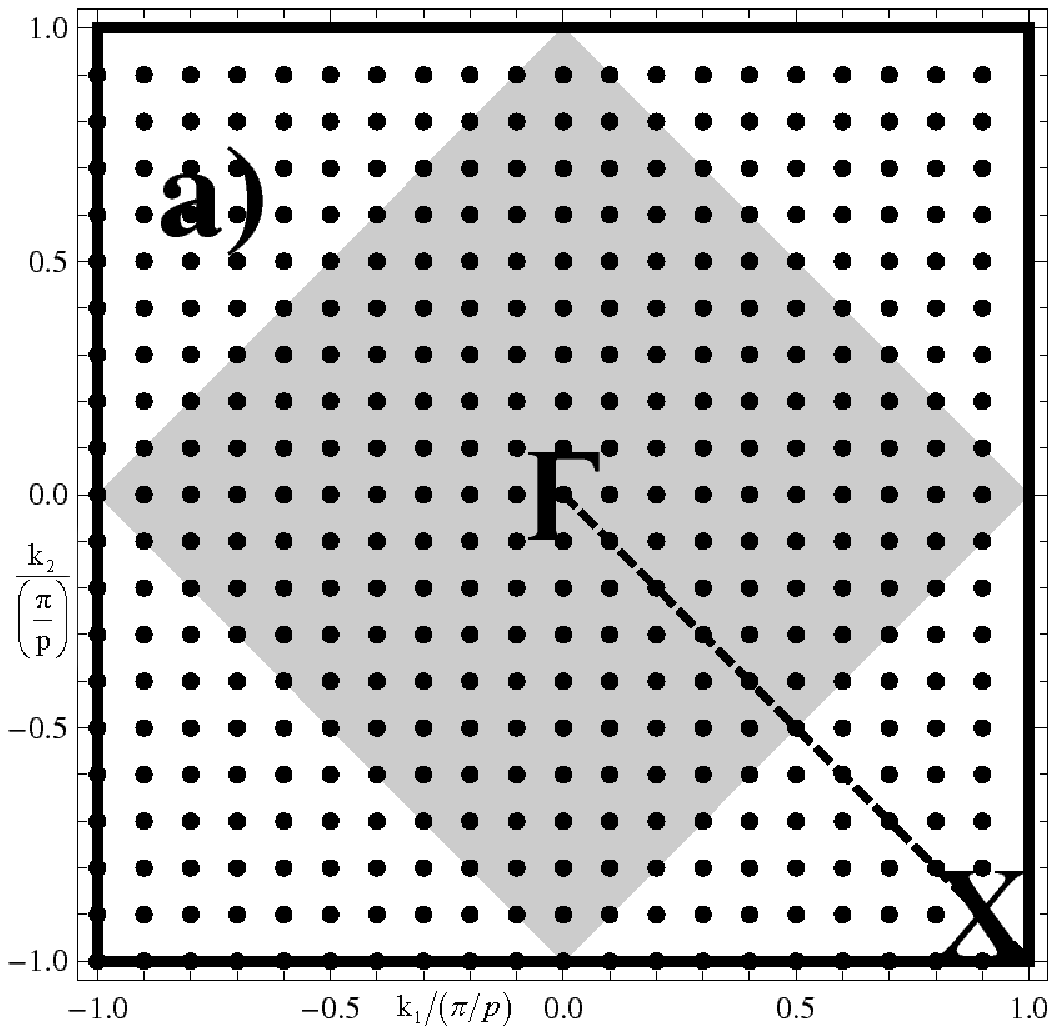}
\ \
\includegraphics[width=1.5in,height=1.5in]{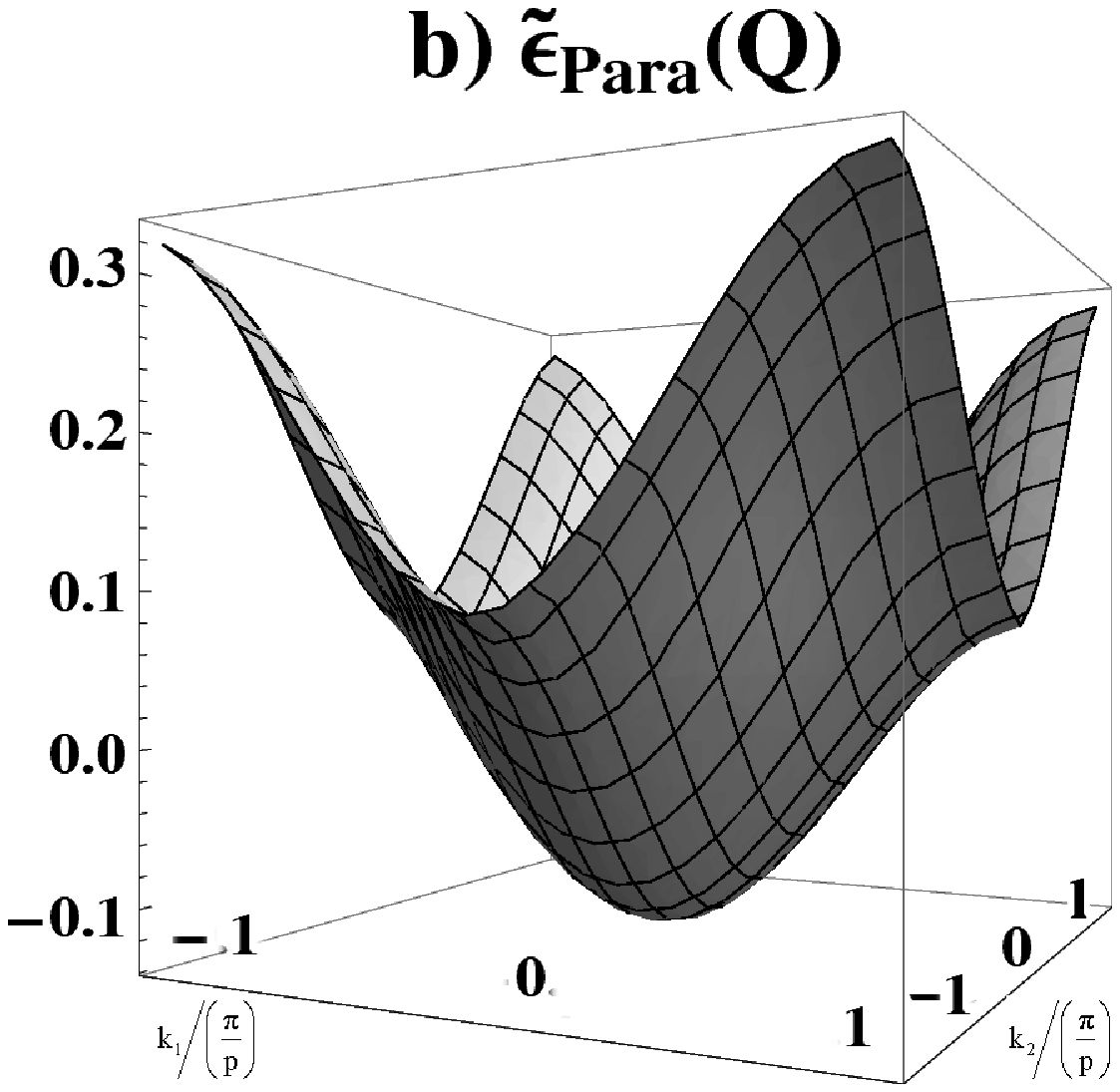}
\end{center}
\caption{ The figure a) shows  the Brillouin zone associated to the
absolute point lattice. The grey zone signals  the occupied states
in the paramagnetic metallic solution at half filling conditions.
The unity of quasimomentum is $\frac{\pi}{p}$. Figure b) shows the
doubly degenerated bands associated to the same paramagnetic and
metallic state. Note the close correspondence between these results
and the ones obtained by Matheiss in Ref. \onlinecite{matheiss}. The
zero energy level in all the band diagrams is the Fermi energy of
the isolator and antiferromagnetic solution presented in subsection
\ref{solAntiFerro}. The domain of the plot is the B.Z. of the
sublattice shown in Fig. 2 a).
 }\label{f:bM}
\end{figure}

Analogously to the ones showed in the previous section, in this
representation, the normalization condition for the single particle
states and the  HF energy of the system take the forms
\begin{widetext}
\begin{eqnarray}
1&=&\bar{\textrm{B}}^{\textbf{Q},l^*}.\bar{\textrm{I}}_{\textbf{Q}}.
\bar{\textrm{B}}^{\textbf{Q},l},\\
\bar{E}^{HF}_{\textbf{Q},l}&=&\sum_{\textbf{Q},l}\Theta_{(\widetilde{\varepsilon}_{F}
-\widetilde{\varepsilon}_{l}(\textbf{Q}))}[
\widetilde{\varepsilon}_{l}(\textbf{Q})
-\frac{\widetilde{\chi}}{2}\bar{\textrm{B}}^{\textbf{Q},
l^*}.(\bar{\textrm{G}}^{dir}_{\textbf{Q}}-\bar{\textrm{G}}^{ind}_{\textbf{Q}}).
\bar{\textrm{B}}^{\textbf{Q}, l} ].
\end{eqnarray}
\end{widetext}

The method employed for solving all the self-consistent matrix
problems considered in this work was an iterative one which  started
from a particular and estimated as convenient state configuration.
In the case being examined in this section, for beginning the
iterations, we used a paramagnetic state. The figure \ref{f:bM}
shows the paramagnetic, metallic and doubly degenerate band obtained
from the iterative process on (\ref{EcuMatricial2}). A half filling
condition has been assumed, that is, a state with one electron per
cell is being considered. Specifically, for the case of
$N=20\times20$ electrons, the occupied states inside the B.Z. are
showed in figure \ref{f:bM} a) by the points inside  the shadowed
region. The chosen parameters were: $\epsilon$=10, which is a common
value for semiconductors, $\widetilde{a}$=0.25, $\widetilde{b}$=0.05
and $\widetilde{\gamma}$=-0.03 (see Appendix \ref{AEleM}), by
following the criterium of fixing  a bandwidth of 3.8 eV
\cite{matheiss}. The here obtained band topologically coincides with
the conduction band presented in Ref. \onlinecite{matheiss}. In both
of them the Fermi level on the $\Gamma$-$X$ direction is a square
which vertices touch the middle of the B.Z. (of the CuO lattices)
sides and also the maximal and minimal energies lay on coinciding
points. Therefore, in this subsection the basis of the effective
model employed here and its main set of parameters have been
defined.

\subsection{Insulating and antiferromagnetic solutions.}\label{solAntiFerro}

\begin{figure}[h]
\begin{center}

\includegraphics[width=1.6in,height=1.6in]{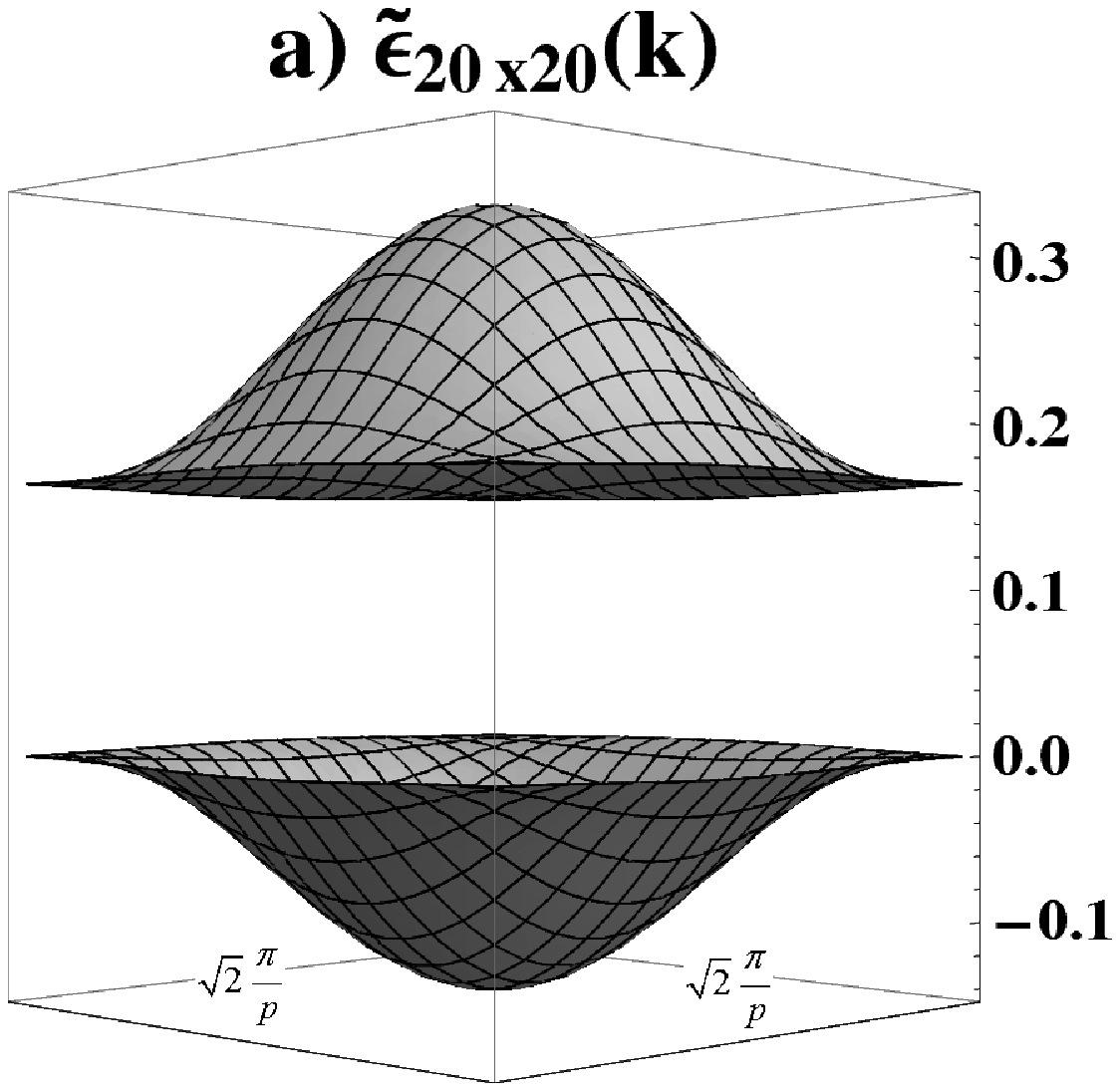}
\ \ \ \
\includegraphics[width=1.6in,height=1.6in]{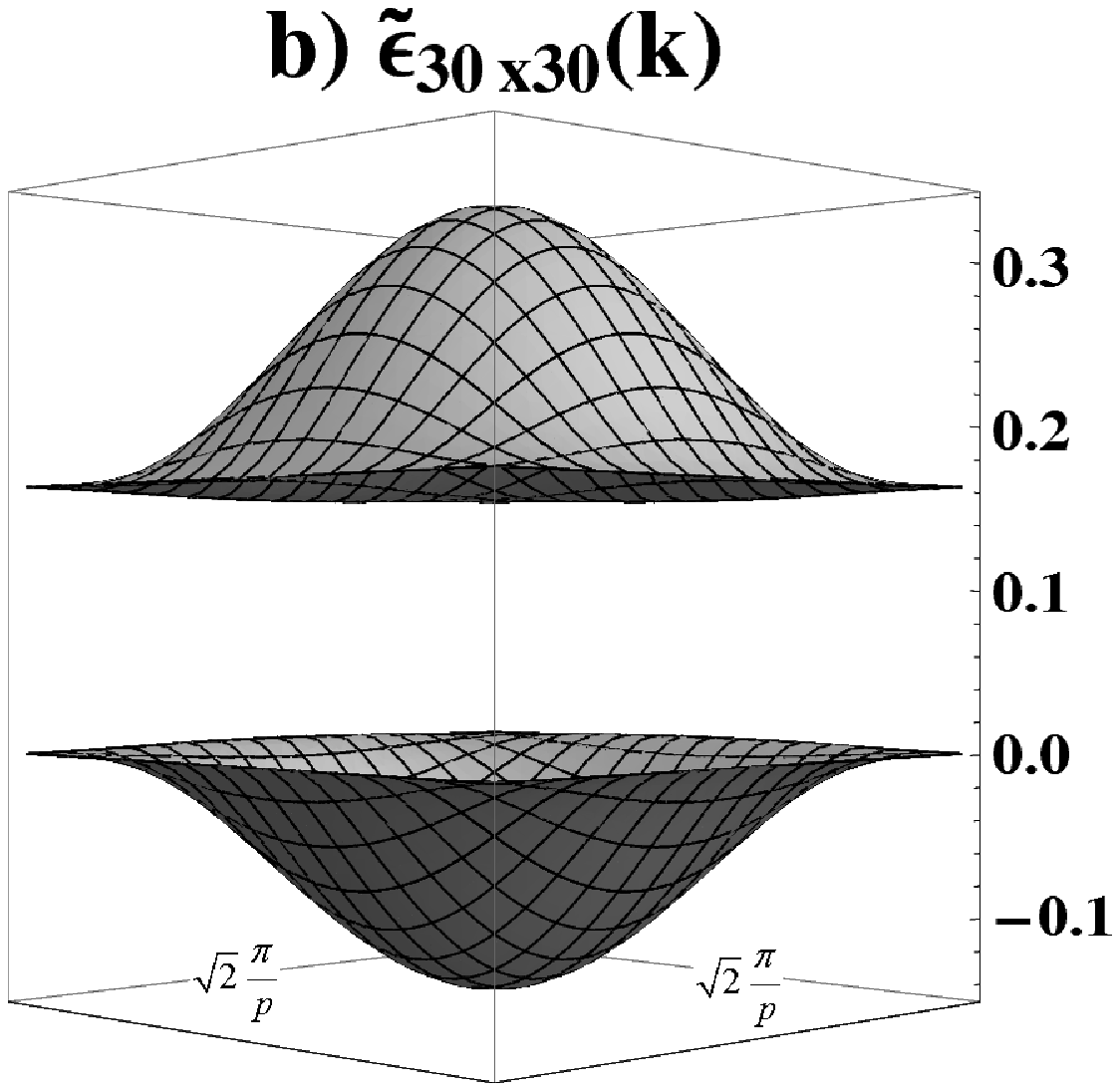}


\end{center} \caption{Energy bands obtained for: a) A sample of  20x20 cells,
 E$_{gap}$ = 1.32 eV. b) A sample of 30x30 cells, E$_{gap}$ = 1.32 eV.
 The parameter values chosen were $\widetilde{a}$ = 0.25,
$\widetilde{b}$ = 0.05, $\widetilde{\gamma}$ = -0.03 and $\epsilon$
= 10. The zero energy level is fixed in the Fermi level of the
20$\times$20 system. Note that the difference in energy between the
two bands is not appreciable in the employed energy scale. The
domains  of both plots is the B.Z. of the sublattice shown in Fig. 2
a) }\label{f:bandas}
\end{figure}

As we have stated before, the solution of the system of equations
Eq. (\ref{EcuMatricial})  was performed  by the method of successive
iterations. The results which presented from now on were found by
employing the parameter values $\epsilon$, $\widetilde{a}$,
$\widetilde{\gamma}$ and $\widetilde{b}$, which were determined in
the previous section. It is important to be noted  that
$\widetilde{a}$, $\widetilde{b}$ $\ll$1 and also that
$\widetilde{\gamma}$ must be of the order of the overlapping among
nearest neighbors factor.  It was necessary to start the iteration
process from a particular state having an antiferromagnetic
character from the beginning,  in order to achieve convergence
toward the solution presented in this subsection. In the figure
\ref{f:bandas} two sets of results following  for a half filling
band are shown. They correspond to two lattices of 20x20 and 30x30
cells. The bands are depicted on the same scale of energies. The
difference between them is of the order of 10$^{-5}$ dimensionless
units of energy $\frac{\hbar^2}{ma^2}$= 8.3 eV. Evidently, they are
bands corresponding to  insulating states.The close similarity of
both results indicates that the thermodynamical limit has being
satisfactorily  achieved for the considered sizes of the periodic
system.
 The HF energy of this HF solution  was  the lowest among  of all the ones found.
  In coincidence  with the experimental evidence, they are states
with local magnetic moment resting on the direction of the
sublattice  x$_{12}$ (see figure \ref{f:magnetizacion}). In the next
section we will show the difference between this HF energy of this
state and the ones corresponding to the other determined HF
solutions and shall comment on this respect.

One important quantity which has been experimentally measured  is
the magnetization.  It is therefore motivating to inspect the
prediction of the obtained  HF state for this magnitude. Its
definition is given by the expression
\begin{eqnarray}\label{e:m}
\textbf{m}(\textbf{x})=\sum_{\textbf{k}^\prime,\textbf{l}}
\sum_{s,s^\prime}\phi^*_{\textbf{k}^\prime,\textbf{l}}(\textbf{x},s)
\boldsymbol{\sigma}(s,s^\prime)\phi_{\textbf{k}^\prime,\textbf{l}}(\textbf{x},s^\prime),
\end{eqnarray}
where
\begin{eqnarray*}
\boldsymbol{\sigma}(s,s^\prime)=\sigma_{x_1}(s,s^\prime)\
\hat{\textbf{e}}_{x_1}+\sigma_{x_2}(s,s^\prime)\
\hat{\textbf{e}}_{x_2}+\sigma_{z}(s,s^\prime)\ \hat{\textbf{e}}_{z},
\end{eqnarray*}
and $\sigma_{x_1}$= $\bigl( \begin{smallmatrix}0 & 1 \\ 1 & 0
\end{smallmatrix} \bigr)$, $\sigma_{x_2}$=$\bigl( \begin{smallmatrix}0 & i \\ -i & 0
\end{smallmatrix} \bigr)$ and $\sigma_{z}$=$\bigl( \begin{smallmatrix}1 & 0 \\ 0 &
-1\end{smallmatrix} \bigr)$ are the Pauli matrices.

In figure \ref{f:magnetizacion} a) the only non vanishing component
of (\ref{e:m}) in this solution is plotted. An interesting result is
that it has been  experimentally observed that La$_2$CuO$_4$ has a
magnetic moment of  0.68 $\mu_B$ per  Cu site on the CuO plane
\cite{pickett}.  The  value obtained  from evaluating the above
formula for our  HF state turns out to be   0.67 $\mu_B$. Therefore,
the considered here discussion  satisfactorily predicts the whole
antiferromagnetic structure of La$_2$CuO$_4$.

\begin{figure}[h]
\begin{center}\label{f:magnetizacion}
\includegraphics[width=1.5in,height=1.5in]{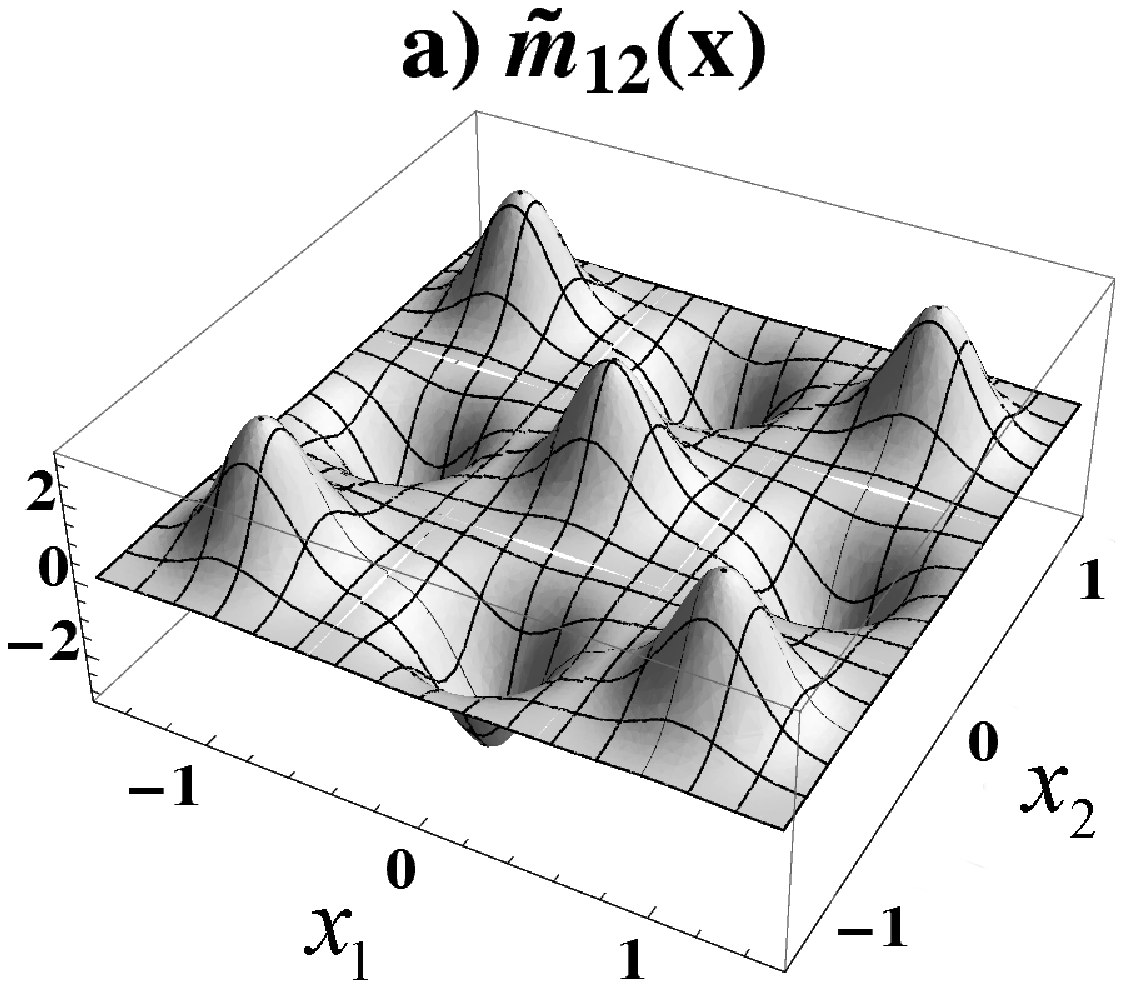}
\ \ \ \
\includegraphics[width=1.5in,height=1.5in]{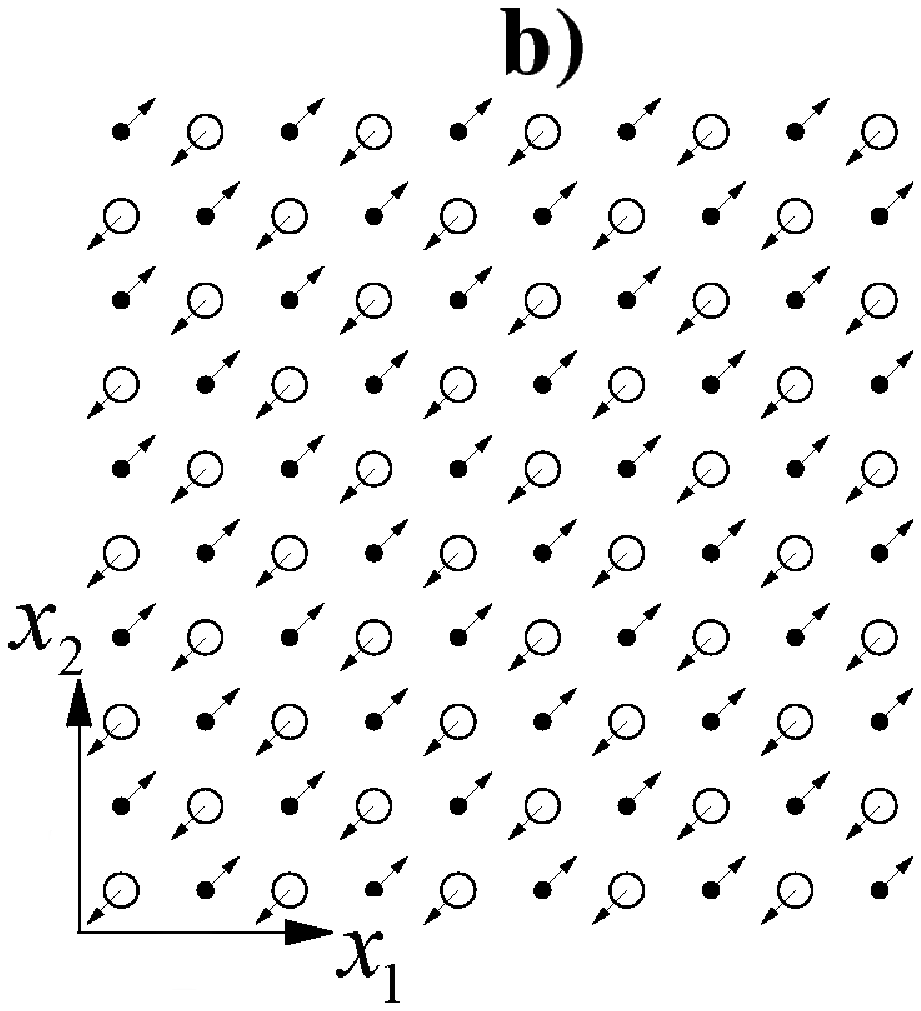}
\end{center} \caption{ The  magnetizaion vector  $\textbf{m}$
of the more stable HF state here determined rests in the direction
1-2. a) This figure shows the projection $\widetilde{m}_{12}$ of the
 dimensionless magnetization, in the  1-2 direction. The magnetization unit is
 $\frac{\mu_B}{p^2}$. b) The picture shows a scheme of the mean magnetic moment per site in the
 lattice. For the shown solution its modular value is 0.67$ \mu_B$.}
\label{f:magnetizacion}%
\end{figure}

It is also helpful to remark that the corresponding single particle
states carry a more intensive antiferromagnetism as more closer they
are from the Fermi surface. Therefore, this property offers a clear
 explanation of the gradual  loss observed in the
antiferromagnetic order under the doping with holes \cite{imada}. In
figure 6  the dependence of the angle $\phi$
 between the magnetic moments per cell on
each of the sublattice 1 and 2, shown by each one of the single
particle Bloch states, is plotted. These components are defined as
the integrals of the magnetic moment density over all the unit cells
of the \textit{absolute} lattice centered in the sublattice points.
Note that the states laying just on the Fermi surface are perfectly
antiferromagnetic ones, and that the more away from the boundaries
the orbitals are,  the less antiferromagnetic they become. Then,
this HF solution indicates that after the orbitals are allowed to
spatial dependent spin orientations, the electrons prefer to
reorient their spin when traveling between contiguous lattice cells.
This effect that can be interpreted as
 clean "correlation" effect  when considered within a restricted HF picture..

\begin{figure}[h]
\begin{center}
\includegraphics[scale=0.5]{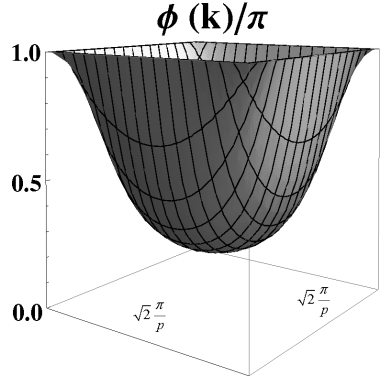}
\caption{The single particle states exhibit  a sharp
antiferromagnetism in the proximities of B.Z. boundaries. In the
figure the angle between their  magnetic moment components on each
of both sublattices (after divided by $\pi$) is plotted against
their Bloch states quasimomenta. Note that the states  on the
boundary have a perfect antiferromagnetism and that they become less
antiferromagnetic as  their quasimomenta move away from the
boundary.  The region of the plot is the B.Z. of the sublattice
shown in Fig. 2 a) }
\end{center}
\label{f:angulo}
\end{figure}
It  also follows  that the size of the zone in which the
antiferromagnetism is strongest,  inversely depends on the
dielectric constant. Thus, the less is the Coulomb interaction among
the electrons in the half filled electron  band,  smaller becomes
the antiferromagnetic character of the single particle states and
the region in which the antiferromagnetism accumulate. Only the
single particle states staying exactly on the Fermi surface  remain
having a perfect antiferromagnetism. In addition, the magnitude of
the gap also decreases with the increasing of the dielectric
constant.

\subsection{Paramagnetic solution showing a pseudogap.}\label{solPara}

It can be observed that to remove the translational symmetry
restrictions does not mean renouncing to obtain a paramagnetic state
as solution. As a matter of fact, this solution effectively exists
and its properties are quite interesting. In Section \ref{sec1:1} we
had already commented about La$_2$CuO$_4$ and its antiferromagnetic
{\em Mott's insulator} properties. However, at an intermediate level
of doping, this material presents very special properties. After the
breaking of the  antiferromagnetic order and in a certain
temperature T  and doping ranges, the material transits to phases
showing a so called  pseudogap. In the La$_2$CuO$_4$ and other  HTc
materials the
 presence of this property  has been observed in some regions
of their  paramagnetic phase \textbf{PM} and in the  superconductor
phase \textbf{SC} \cite{vonharlingen}.

\begin{figure}[h]
\begin{center}\label{f:bandas1}

\includegraphics[width=1.5in,height=1.5in]{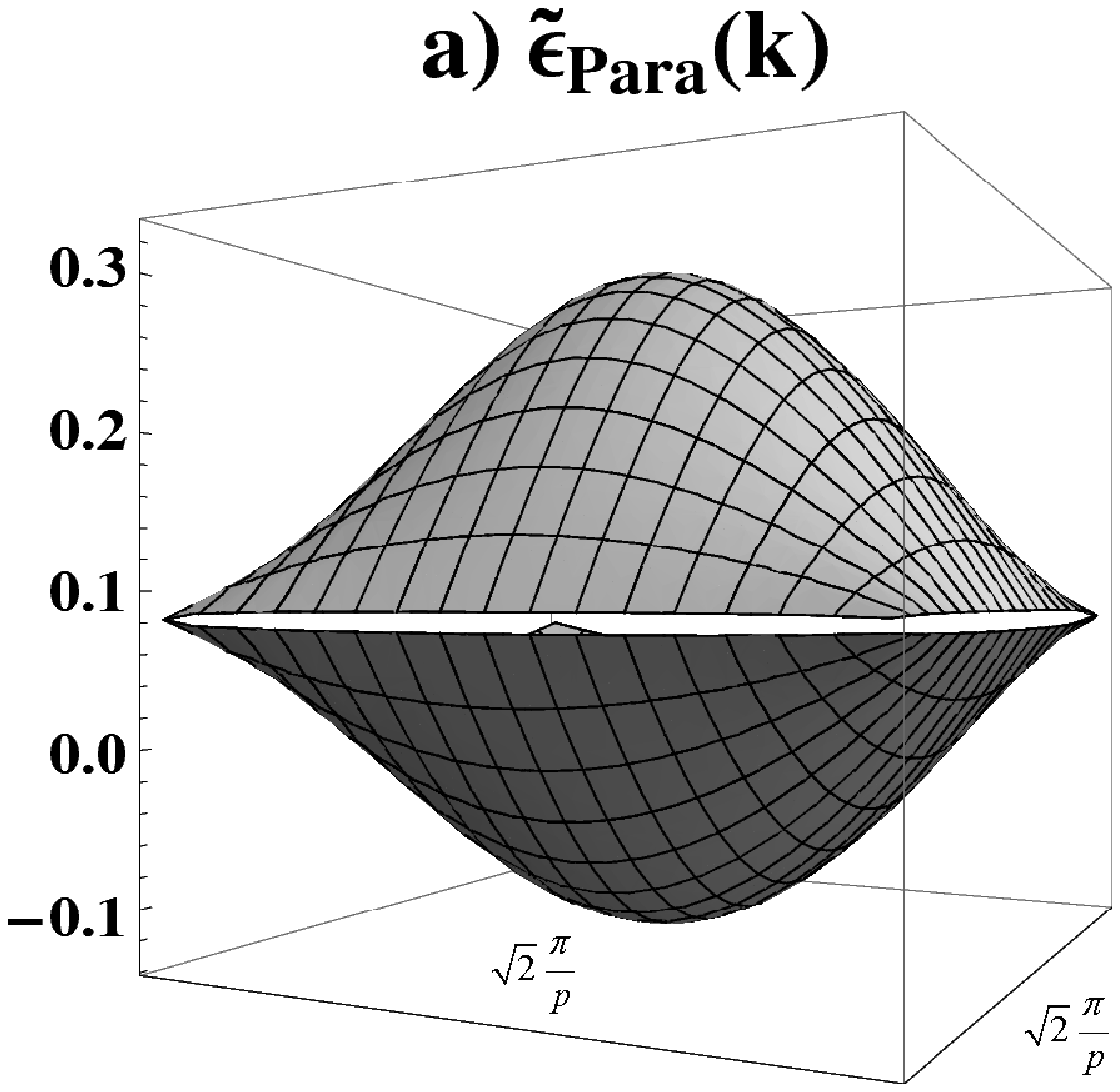}
\ \ \ \
\includegraphics[width=1.5in,height=1.5in]{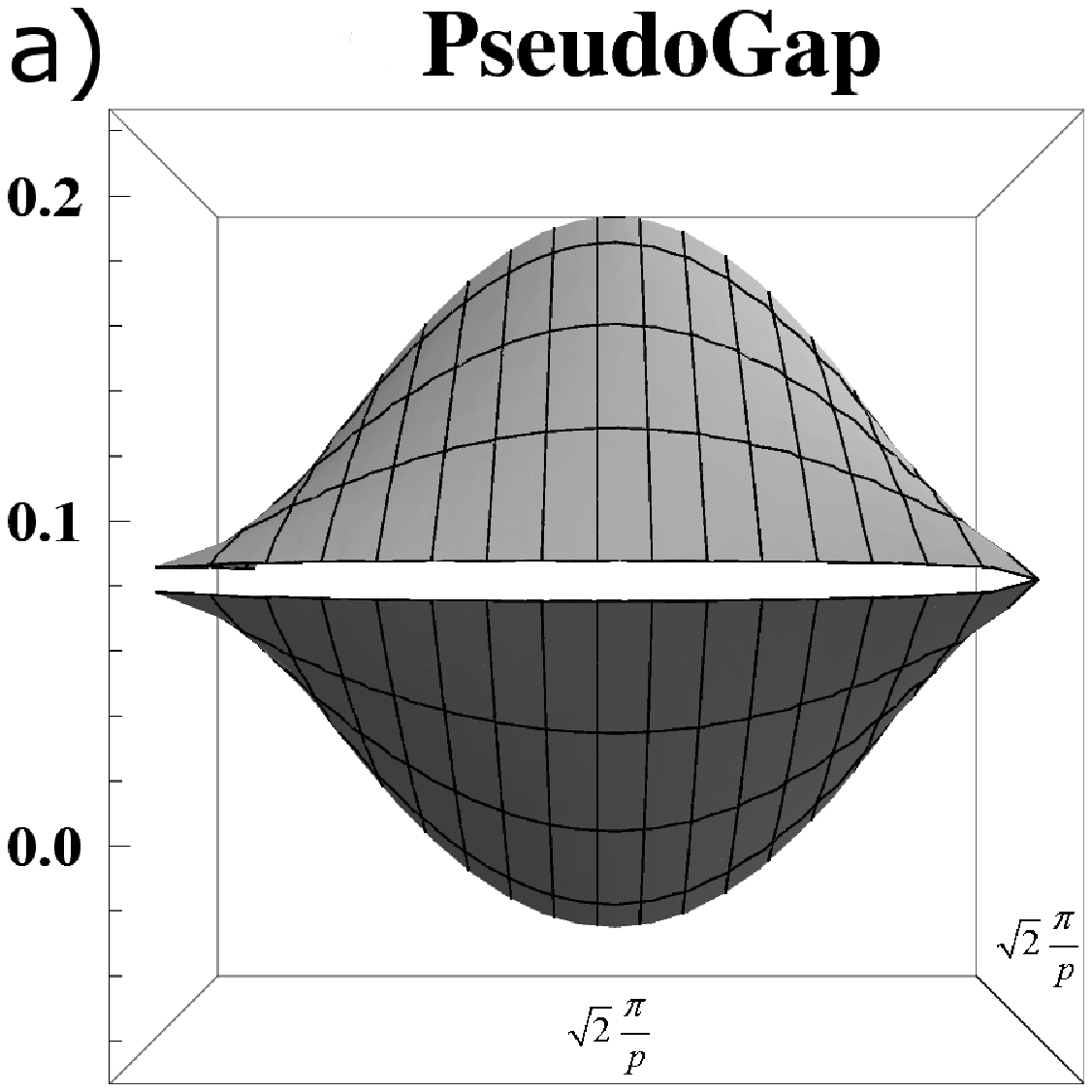}
\end{center} \caption{In the figure: a)\ The band structure associated to
 paramagnetic ground state showing a pseudogap is shown. b)\
A frontal view of the plot more clearly showing the pseudogap. The
zero energy level is the Fermi energy of the IAF solution. The two
graphics are plotted in the B. Z. of the sublattices, that is,  the
grey zone in figure 2 a).}
\label{f:bandas1}%
\end{figure}
In figure \ref{f:bandas1}  the band  spectra corresponding to the HF
paramagnetic ground state obtained from Eq. (\ref{EcuMatricial}) is
shown. Note the existence of a pseudogap which reaches a maximum
value of 0.1 eV $\approx$ 10$^3$ K (equivalent to 0.012
dimensionless unit of energy $\frac{\hbar^2}{ma^2}$ = 8.3 eV). The
parameters given in the Subsection \ref{SMST} were employed for this
evaluation. The pseudogap amplitude rises for smaller dielectric
constants $\epsilon$, that is, with the reduction of the screening.
It is an interesting result that  the  HF energy of this ground
state is exactly coincident with the one corresponding to the
paramagnetic and metallic state presented in Subsection \ref{SMST}.
Moreover, the lowest energy band  in both solutions, also exactly
coincide, with an  upper bound error of $10^{-6}$ in dimensionless
energy units (that means $10^{-5}$ in eV). Thus, the occupied single
particle states in both solutions are identical and in consequence
the momentum dependence of the filled energy  bands also coincide.
Henceforth,  the difference between the two solutions refers only to
the non occupied states. Those states can not be considered in the
band model presented in Subsection \ref{SMST}. For instance,
consider the expressions for two states, one occupied an another
empty, which are associated to the same four quasimomentum value
\begin{widetext}
\begin{eqnarray}
\textrm{B}^{\textbf{K}}_o= \begin{pmatrix} \frac{1}{\sqrt{2}}\\
\frac{1}{\sqrt{2}} \\0\\0\end{pmatrix} \longrightarrow &
\phi_o(\textbf{x},s)=\sqrt{\frac{1}{N}}\ u_{\uparrow}(s)
\sum_{r,\textbf{R}^{(r)}} \ \exp(i\
\textbf{K}\cdot\textbf{R}^{(r)})\
\varphi_{\textbf{R}^{(r)}}(\textbf{x}), \label{ocupado} \\
\textrm{B}^{\textbf{K}}_e=\begin{pmatrix} \frac{1}{\sqrt{2}}
\\- \frac{1}{\sqrt{2}}\\0\\0\end{pmatrix}\longrightarrow & \phi_e(\textbf{x},s)=\sqrt{\frac{1}{N}}\
u_{\uparrow}(s)\sum_{r,\textbf{R}^{(r)}}\ (-1)^r \exp(i\
\textbf{K}\cdot\textbf{R}^{(r)})\
\varphi_{\textbf{R}^{(r)}}(\textbf{x}), \label{exitado}
\end{eqnarray}
\end{widetext}
where $\varphi_{\textbf{R}^{(r)}}$ represents the gaussian orbitals
$\varphi_{0}$ centered on $\textbf{R}^{(r)}$.

Comparing with the expression (\ref{estadoPMatheiss}),  it may be
noted that the occupied single particle state (\ref{ocupado}) can be
expressed on the basis (\ref{baseMatheiss}), and in fact, coincides
with the state $\bar{\phi}^{\ \textbf{K},\uparrow}_o$ obtained from
solving (\ref{EcuMatricial2}). However the excited state
(\ref{exitado}) was  not allowed  in the space of orbitals employed
in solving  (\ref{EcuMatricial2}), and thus its attainment  is a
neat consequence of the removing of the usually imposed symmetry
under the maximal group of translations. More precisely, these
excited states showing a pseudogap, appeared thanks to the allowed
independence between the Bloch functions defined in both
sublattices,  introduced in this work. This result shows an
interesting potential use that this kind of constraints removing
procedure  could introduce  in the determination of new excitation
properties of the HF solutions in various applications.

In the following table  the HF energies per particle are shown for
the paramagnetic-metallic (PM), paramagnetic with pseudogap (PPG)
and insulator-antiferromagnetic (IAF) ground states; with zero point
energy assumed on the last one:

\begin{table}[h]

\begin{tabular}{||r||r||r||r||}
\hline\hline State & IAF & PM & PPG \\
\hline\hline $\Delta$E (eV) & 0.0 & +0.076 & +0.076 \\
\hline\hline
\end{tabular}
\caption{Energy differences  between the various HF states.}
\label{table1}
\end{table}

It is an interesting  outcome that the energy difference PM
(PPG)-IAF and the N\'eel temperature of this kind of  materials are
both of the order of 10$^2$ K. Thus,  the results suggest the
possibility of
 having further success in applying the approach started in this work to
the description of some regions of the phase diagram of the La$_2$
CuO$_4$.  On the question about the relative stability between the
PM and PPG states, it can be estimated  that due to the presence of
a pseudogap, the PPG ground state should become more stable than the
PM at  non vanishing temperatures. This should be the case because
for creating excitations on the PPG ground state, temperatures of
hundreds of Kelvin degrees are needed, while excitations in the PM
ground state can occur in every range of temperatures.

\begin{figure}[h]
\begin{center}\label{f:bandas2}
\includegraphics[width=2in,height=2in]{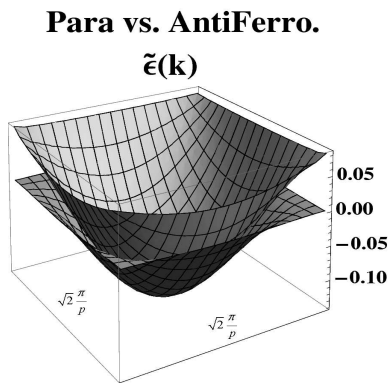}
\end{center} \caption{The figure shows in the same plot the occupied bands
corresponding to the states  PPG and IAF. The difference in the
energies of the orbitals is concentrated in the boundary of the
Brillouin zone. The zero energy level coincides with the Fermi level
of the IAF solution. The domain of the plot is the B.Z. of the
sublattices, given by the grey zone in figure 2 a). }
\label{f:bandas2}%
\end{figure}

In figure \ref{f:bandas2} the PPG (PM) and IAF occupied bands are
depicted  in a common frame. The main difference in their energies
corresponds to the single particle states being closer to Fermi
surface. As we had noted  before, the same behavior has the
antiferromagnetic character of the  single particle states of the
IAF solution. Therefore, it can be hoped that from the solution of
(\ref{EcuMatricial2}) under doping with holes, both solutions move
toward a common ground state without absolute magnetic order, in
correspondence with the pattern shown by the phase diagrams of these
materials \cite{pickett}. That is, at a critical doping level value
 a breaking of the antiferromagnetic order could occur  and
 the system presumably transits to a phase showing  a pseudogap.

\bigskip
\bigskip

\section{Conclusions} \label{chapter5}

The results of this  work support the potential of  the HF
self-consistent method  for the description of some secular
properties exhibited by materials such as copper-oxygen compounds,
specifically La$_2$CuO$_4$. Those are properties usually associated
to strong correlation effects, that could  be explained by {\em
first principle} calculations,  after  modified to incorporate
procedures to search for  spontaneous symmetry breaking solutions
and general  ways of exploring  the spin structure of the HF
obitals. Note that single particle orbitals are not only employed in
the HF method. They are also essential ingredients of more general
schemes as the various types of density functional methods. Thus,
the kind of reasoning employed in this work seems to be easily
easily implemented in such discussions. In the present work we
explore those paths in a simplified manner. In order to avoid the
intrinsic complexities of La$_2$CuO$_4$ material, it was necessary
to utilize a simple model which was sufficiently flexible for
reproducing the dispersion profile of the only half filled
  band  of La$_2$CuO$_4$ reported in Ref. \onlinecite{matheiss}.
  After imposing the maximal symmetry under
translations a paramagnetic and metallic ground state (PM) was
obtained  which dispersion properties topologically also coincide
with the results given in Ref. \onlinecite{matheiss}. Then, the free
parameters of effective model  were fixed from the requirement
 of reproducing the bandwidth of the half filled band obtained in Ref.
\onlinecite{matheiss}. Employing those parameters and from removing
some symmetry restrictions, solutions  were obtained which show
interesting properties. In agreement with the experiment, the
insulating-antiferromagnetic (IAF) solution turns to be the most
stable among all HF states found.  Some of its properties are
enumerated below:
\begin{enumerate}
\item The gap magnitude diminish with the increasing of the
screening constant $\epsilon$.
\item The antiferromagnetic structure the HF orbitals increases
when the states approach the Fermi level. In addition, the size of
the outlying region in which antiferromagnetism persists depends on
the screening created by the effective environment $\epsilon$. That
is, by increasing  screening, the  size of the antiferromagnetic
region reduces. Thus,  the idea arises that after
 doping with holes (that is, solving for HF solutions not at half filling
 condition as it is done here)   the antiferromagnetic zone, which is precisely
 concentrated near the Fermi level could be annihilated, producing in this way
 a phase transition to a non  magnetically ordered
 ground state. This possibility indicates a way to describe the normal state
 properties of the HTc superconductors through a simple HF study.
\item The magnetic moments per cell which are evaluated  rest on the right
direction respect to the lattice and its modular value of 0.67
$\mu_B$ satisfactorily coincides with the experimentally measured
result 0.68 $\mu_B$. This outcome support the adjust made of the
parameters defining  the effective model employed.
\end{enumerate}

The other HF solution presented in this work corresponds to a
paramagnetic state showing a pseudogap (PPG). Few  properties of
this state are described in what follows:
\begin{enumerate}
\item The pseudogap magnitude diminishes with the increasing of the
screening constant $\epsilon$.
\item Just as we hoped from the beginning of the work, at zero temperature the
states PM and PPG resulted identical. The difference between them is
only given by the excitations of the system. Thus, the removal of
some symmetry restrictions, defines in this case, new properties for
unoccupied single particle states. It seems feasible that in other
materials,  it could be possible to obtain a gap instead of a
pseudogap even in the absence  of magnetic order. Such an outcome
could show the ability of a properly formulated HF description in
describing all the kinds of Mott insulators, being or not
magnetically ordered.
\item The amplitude of the obtained pseudogap reaches a value of the
order of 10$^2$K. Thus, it is possible that the system  at finite
temperature  should prefer the PPG state over the PM one. That is an
interesting signal, knowing that La$_{2-x}$M$_x$CuO$_4$ presents
pseudogap in the \textbf{PM} phase.
\item In the same way as it happens for the antiferromagnetic character of the IAF
ground state, the difference between the one particle energies of
IAF  and the PPG (PM) states, is greater for states being closer to
the Fermi surface. That is, it happens for the electrons with more
energy and consequently the first ones in disappear under doping
with holes. Therefore, as it was mentioned before, these results
suggest the possibility to describe a transition IAF$\rightarrow$PPG
under doping with holes, within the here studied effective model.
\end{enumerate}

 One methodological conclusion of this work seems to be worth to be mention.
 It corresponds to  the fact that
 the results  fully clarifies  that  the solutions of a general HF problem
 non necessarily should turn to be a set of single particle
 orbitals,  all having an $\alpha$ or $ \beta$ spin structure at all  points of the space.
 This is directly shown by the  particular example of the IAF solution in
 which the HF single particle states have a neatly non separable
 character in their spin and orbital dependence. This possibility,
 although being simple, and perhaps even expected to be correct
 after to direct the attention to it,  looks to be  relevant
  for the description of magnetic and strongly correlated systems.

The above enumerated conclusions motivated  new objectives to be
considered in extending the work. Our general interests are twofold:
in one side to achieve a more complete and precise investigation
about the here detected unexplored potentialities of the {\em first
principle} calculations. In another direction,  to look for the
possibility of describing the HTc superconductivity  in the
framework of the here studied simple model.  Some of the more
specific issues of future searches in connection those general
objectives are the following ones:
\begin{itemize}

\item To generalize the discussion done in this work to introduce the
 doping with holes as a new parameter. Then, it will be
possible to investigate the effects of the doping on  the here
determined ground states. Of particular relevance in this sense
appears the IAF state.
\item To compute the zero temperature electron Green function of the system,
and use  it to evaluate the effective polarization of the
La$_2$CuO$_4$ in the obtained states.
\item With the polarization results at hand, attempt to solve the Bethe-Salpeter
equation for two holes in the IAF ground state (and also in the PPG
one), to find whether it is possible or not to define the existence
of preformed Cooper Pairs in the model. The possibility for its
existence was suggested by the results of Ref.
\onlinecite{caboponce}, in which it was argued that a strong
2D-screening of the Coulomb interaction is created by a half filled
band of tight binding electron.
\item To continue the application of the ideas advanced here
in help clarifying  the widely known  debate between the Mott and
Slater pictures  about the electronic structure of solids.
\item To apply a rotational invariant HF calculation for obtaining
states of molecular or atomic systems   having incomplete shells. We
guess, that this application could  help to reduce the value of the
correlation energies in those systems. This idea is suggested by the
results of this work which
 show that the HF energy can be optimized in many of the currently considered
 physical systems, due to the difficulties in determining the best among
 all possible self-consistent solutions.
\end{itemize}

\begin{acknowledgments}
We express our gratitude to A. Gonz\'alez, C. Rodr\'iguez, A.
Delgado, Y. Vazquez-Ponce and N. G. Cabo-Bizet by  helpful
conversations and comments. In addition, the relevant support
received from the Proyecto Nacional de Ciencias B\'asics (PNCB,
CITMA, Cuba) and from the Network N-35 of the Office of External
Activities (OEA) of the ASICTP (Italy) are deeply acknowledged.
\end{acknowledgments}
\bigskip

\appendix
\section{Matrix elements.}\label{AEleM}

\subsection{Brackets notation}\label{sub:Brac}
The bracket terms in (\ref{energiat}) represent the following
integrals:
\begin{widetext}
\begin{eqnarray}
\\\nonumber\langle m|\hat{\h}_{0}|p \rangle&\equiv&\sum_s\int d^2x
\ \phi^*_{m}(x,s)\ \hat{\h}_0(x) \ \phi_{p}(x,s),\\\nonumber \langle
m,n|V|o,p\rangle&\equiv&\sum_{s,s^{\prime}}\int d^2xd^2x^{\prime}\
\phi^*_{m}(x,s)\phi^*_{n}(x^{\prime},s^{\prime})\ V(x,x^{\prime})\
\phi_{o}(x^{\prime},s^{\prime})\phi_{p}(x,s),
\end{eqnarray}
\end{widetext}
where m, n, o and  p,  denote anyone of the possible quantum number
arrays.

\subsection{Dimensionless definitions.}
Before writing the expressions for  the  operators matrix elements,
we begin by showing some other useful definitions, as for instance
the employed dimensionless ones:
\begin{widetext}
\begin{eqnarray}\label{e:dimensiones}
\widetilde{p}&\equiv&1 \hspace{4.7cm} \text{ Unit of
distance},\\\vspace{0.5cm}
\widetilde{a}&\equiv&\frac{a}{p} \hspace{3cm} \text{Dimensionless characteristic length},\\
\widetilde{\textbf{R}}&\equiv&\frac{\textbf{R}}{p} \hspace{2.9cm}
\text{Dimensionless lattice points position},\\
\widetilde{V}&\equiv&\frac{ma^2}{\hbar^2}\ V \hspace{2.3cm} \text{
Dimensionless Coulomb potential}.
\end{eqnarray}
\end{widetext}
From those definitions,  the used reference conventions can be
inferred. That is, the distances are expressed in units of the Cu
nearest neighbors separation {\em p} = 3.6 \AA; and the energies and
potential interactions, in units of  the quantity
$\frac{\hbar^2}{ma^2}$, that for the employed parameters is
equivalent to 8.3 eV.

\subsection{Other definitions and properties.}
 We had already defined along the paper the Wannier orbitals
$\varphi_{\textbf{R}^{(r)}}$, as normalized gaussian functions,
centered in $\textbf{R}^{(r)}$ and with characteristic parameter
{\em a}. In order to simplify notation when the jellium in
(\ref{freehamiltonian}) is going to be used, is useful to define:
\begin{eqnarray}\label{Wannierfondo}
\varphi^{b}_0(\textbf{x})=\frac{1}{\sqrt{\pi
b^2}}\exp(-\frac{\textbf{x}^2}{2\ b^2}).
\end{eqnarray}
Let us use (\ref{Wannierfondo}) for defining the jellium's potential
\begin{eqnarray}\label{fondoapendice}
F_b(\textbf{x})&=&\sum_{\textbf{R}}\int d^2y\
\varphi^{b^*}_{\textbf{R}}(\textbf{y})V(\textbf{x}-\textbf{y})
\varphi^{b}_{\textbf{R}}(\textbf{y}).
\end{eqnarray}
In obtaining the matrix elements of the jellium potential
(\ref{fondoapendice}), the following notation is useful:
\begin{widetext}
\begin{eqnarray}\label{notacionint}
\langle
\textbf{R}^{(r)},\textbf{R}^{b}|V|\textbf{R}^{b},\textbf{R}^{(t)}\rangle&\equiv&\int
d^2xd^2y\
\varphi^*_{\textbf{R}^{(r)}}(\textbf{x})\varphi^{b^*}_{\textbf{R}}(\textbf{y})V(\textbf{x}-\textbf{y})
\varphi^{b}_{\textbf{R}}(\textbf{y})\varphi_{\textbf{R}^{(t)}}(\textbf{x}),
\end{eqnarray}
\end{widetext}
where labels {\em r} and {\em t} move on the sublattices
independently. In the context of the infinite lattice problem is
easy to see the following property
\begin{eqnarray}\label{integralfondo}
\nonumber\langle\textbf{R}^{(r)},\textbf{R}^{b}|V|\textbf{R}^{b},\textbf{R}^{(t)}\rangle&=&
\langle\textbf{R}^{(r)}-\textbf{R},\textbf{0}^{b}|V|\textbf{0}^{b},\textbf{R}^{(t)}
-\textbf{R}\rangle, \\
\end{eqnarray}
which can be proved by making a pair of changes of variables on the
second order integral in the right hand side of (\ref{notacionint}).
For finite systems the prove is a little more complicated. Firstly,
the integrals extend over the region occupied by the lattice, and
for this region the integration does not remain invariant under the
translations that must be done. It becomes necessary to periodically
extend the Coulomb interaction beyond the boundaries. That is,
modified it to the form
 \begin{equation}
V_p(\textbf{x},\textbf{y})=\frac{e^2}{4\pi\epsilon\epsilon_0}
\sum_{n_{1},n_{2}}\frac{1}{|\textbf{x}-\textbf{y}+n_{1} L \
\hat{\textbf{e}}_{x_1}+n_{2} L\ \hat{\textbf{e}}_{x_2}|},
\end{equation}
where $n_{1}$ and $n_{2}$ = ..., -1, 0, 1, ..., and $L$ is the
length of the side of the squared region which the system occupies.
However when the system is sufficiently large,  the error done  by
no extending it periodically  is not important; and it vanishes in
thermodynamic limit.

Equally useful in defining the direct and exchange potential matrix
elements, is the following notation
\begin{widetext}
\begin{eqnarray}
\langle
\textbf{R}^{(r)},\textbf{R}^{(t^{\prime})}|V|\textbf{R}^{(t^{\prime\prime})},\textbf{R}^{(t)}\rangle&\equiv&\int
d^2xd^2y\
\varphi^*_{\textbf{R}^{(r)}}(\textbf{x})\varphi^*_{\textbf{R}^{(t^{\prime})}}(\textbf{y})V(\textbf{x}-\textbf{y})
\varphi_{\textbf{R}^{(t^{\prime\prime})}}(\textbf{y})\varphi_{\textbf{R}^{(t)}}(\textbf{x}),
\end{eqnarray}
where $t^{\prime}$ and $t^{\prime\prime}$ also move on both
sublattices independently. In the same manner they fulfill  the
following property
\begin{eqnarray}\label{propdir}
\langle\textbf{R}^{(r)},\textbf{R}^{(t^{\prime})}|V|
\textbf{R}^{(t^{\prime\prime})},\textbf{R}^{(t)}\rangle&\equiv&\langle
\textbf{R}^{(r)}-\textbf{R}^{(t^{\prime\prime})},\textbf{R}^{(t^{\prime})}-
\textbf{R}^{(t^{\prime\prime})}|V|\textbf{0},\textbf{R}^{(t)}-
\textbf{R}^{(t^{\prime\prime})}\rangle,
\end{eqnarray}
\end{widetext}
if we define
\begin{eqnarray}
\textbf{R}^{(t^{\prime},t^{\prime\prime})}&\equiv&\textbf{R}^{(t^{\prime})}-
\textbf{R}^{(t^{\prime\prime})}.
\end{eqnarray}
Then it follows
\begin{eqnarray}
\textbf{R}^{(t^{\prime},t^{\prime\prime})}&=&\begin{cases}\textbf{R}^{(1)}, & \text{ if\ t$^{\prime}$=t$^{\prime\prime}$},\\
\textbf{R}^{(2)}, & \text{ if
t$^{\prime}\neq$t$^{\prime\prime}$}.\end{cases}
\end{eqnarray}
Let us consider the definitions of \textbf{R}$^{(1)}$ and
\textbf{R}$^{(2)}$ given in (\ref{e:subred}), and recall that the
only Wannier orbitals which have non vanishing overlapping are those
centered on the same site or the ones centered in nearest neighbors
(the closest neighbors belong to different sublattice). Then for
fixed \textbf{R}$^{(r)}$ and \textbf{R}$^{(t)}$, the only no
vanishing among all the quantities in the left hand side of
(\ref{propdir})  are
\begin{eqnarray}\label{integral1}
\\\nonumber\langle\textbf{R}^{(r)}-\textbf{R}^{(t^{\prime\prime})},\textbf{0}|V|\textbf{0},\textbf{R}^{(t)}-\textbf{R}^{(t^{\prime\prime})}\rangle
\ \ \text{ for t$^{\prime}$=t$^{\prime\prime}$,}
\end{eqnarray}
and
\begin{eqnarray}\label{integral2}
\\\nonumber\langle\textbf{R}^{(r)}-\textbf{R}^{(t^{\prime\prime})},\textbf{p}_i|V|\textbf{0},\textbf{R}^{(t)}-\textbf{R}^{(t^{\prime\prime})}\rangle
\ \ \text{ for t$^{\prime}\neq$t$^{\prime\prime}$,}
\end{eqnarray}
in which {\em i}=1,...,4. The four quantities \textbf{p}$_i$ are
defined as
\begin{eqnarray}
\textbf{p}_i&=&\begin{cases}p\ \hat{\textbf{e}}_{x_1}, & \text{ if\ {\em i}=1},\\
-p\ \hat{\textbf{e}}_{x_1}, & \text{ if\ {\em i}=2},\\ p\
\hat{\textbf{e}}_{x_2}, & \text{ if\ {\em i}=3},\\ -p\
\hat{\textbf{e}}_{x_2}, & \text{ if\ {\em i}=4}.\end{cases}
\end{eqnarray}
That is, they move over the neighbors which are closest to the site
on the origin.

The employed  procedure reduces the number of integrals appearing in
(\ref{integral1}), (\ref{integral2}), inclusive those corresponding
to the right hand side in (\ref{integralfondo}) and anyone of those
appearing when the matrix elements of the periodic potential
W$_{\gamma}$, or the projection of the tight binding Bloch basis
between any two elements,  are searched. By example
\begin{eqnarray}
\langle\textbf{R}^{(r)}|W_{\gamma}|\textbf{R}^{(t)}\rangle
&\equiv&\int d^2x\ \varphi^*_{\textbf{R}^{(r)}}\ W_{\gamma}\
\varphi_{\textbf{R}^{(t)}},\\
\langle\textbf{R}^{(r)}|\textbf{R}^{(t)}\rangle &\equiv&\int d^2x\
\varphi^*_{\textbf{R}^{(r)}}\ \varphi_{\textbf{R}^{(t)}},
\end{eqnarray}
which respectively fulfill the following properties
\begin{eqnarray}
\langle\textbf{R}^{(r)}|W_{\gamma}|\textbf{R}^{(t)}\rangle&=&
\langle\textbf{R}^{(r,t)}|W_{\gamma}|\textbf{0}\rangle,\\
\langle\textbf{R}^{(r)}|\textbf{R}^{(t)}\rangle&=&\langle
\textbf{R}^{(r,t)}|\textbf{0}\rangle,
\end{eqnarray}
given the periodicity of W$_{\gamma}$ in the absolute sublattice.

In the following subsection it is frequently used the symbol
$\delta_{r,t+1}$ in which $t+1$ is not the usual sum of 1, but he
transformation of a given sublattice in another
\begin{eqnarray}
t+1&=&\begin{cases} 2&\text{  if t=1},\\
1&\text{  if t=2}.
\end{cases}
\end{eqnarray}

\subsection{ Matrix Elements.}\label{annex:matricialelements}
 Making use of the  previously definitions given in this Appendix and
after performing  an extensive algebraic work, the desired matrix
elements are computed. Below we start presenting them:
\begin{widetext}
\begin{eqnarray}
E^0_{\textbf{k},(t,\alpha_z),(r,\sigma_z) }&=&\delta_{
\alpha_z,\sigma_z} \left[\widetilde{W}_{00}\
\delta_{t,r}+2\widetilde{\gamma}(\cos k_1p+\cos k_2p)\
\delta_{t,r+1}\right],\\\nonumber&&
\\ I_{\textbf{k},(t,\alpha_z),(r,\sigma_z)}&=&\delta_{
\alpha_z,\sigma_z} \left[I_{00}\ \delta_{t,r}+2I_{01}(\cos k_1p+\cos
k_2p)\ \delta_{t,r+1}\right],
\\\nonumber&& \\ F_{\textbf{k},(t,\alpha_z),(r,\sigma_z) }&=&\delta_{
\alpha_z,\sigma_z} \left[F_{00}\ \delta_{t,r}+2F_{01}(\cos k_1p+\cos
k_2p)\ \delta_{t,r+1}\right],\\\nonumber
\end{eqnarray}\
\end{widetext}
where $\widetilde{W}_{00}$ = 0,  represents a change in the zero
point energy; and $\widetilde{\gamma}$ is a free parameter
describing our lack of knowledge about the periodic potential. The
other appearing parameters are defined as
\begin{eqnarray}\label{e:const1}
I_{00}&=&\langle \textbf{0}|\textbf{0}\rangle\\\nonumber&=&1,
\\\nonumber
\\
I_{01}&=&\langle 0|\widetilde{\textbf{p}}_1 \rangle\\\nonumber &=& \text{ {\Large e}}^{-\frac{1}{4\widetilde{a}^2}}, \\\nonumber \\
F_{00}&=&\frac{2}{N}\sum_{\widetilde{\textbf{R}}}\langle\widetilde{\textbf{R}},\textbf{0}^b|\widetilde{V}|\textbf{0}^b,\widetilde{\textbf{R}}\rangle,\\\nonumber \\
F_{01}&=&\frac{2}{N}
\sum_{\widetilde{\textbf{R}}}\langle\widetilde{\textbf{R}}+\widetilde{\textbf{p}}_1\
,\textbf{0}^b|\widetilde{V}|\textbf{0}^b;\widetilde{\textbf{R}}\rangle,\\\nonumber
\end{eqnarray}
where {\em N} is the number of electrons in the electron  gas and,
as we had already defined, the symbol \, $ \widetilde{}\ $ means
dimensionless.

The matrix elements of the direct potential are
\begin{widetext}
\begin{eqnarray}\label{e:matrizdirecta}
\\\nonumber G^{dir}_{\textbf{k},(t,\alpha_z),(r,\sigma_z)
}&=&\sum_{\textbf{k}^\prime,\textbf{l}}\Theta_{(\varepsilon_{F}-\varepsilon_{\textbf{l}}(\textbf{k}^\prime))}
\delta_{\alpha_{z},\sigma_z}
\times[\delta_{t,r}{B}^{{\textbf{k}^\prime,\textbf{l}}^*}_{(t^\prime,\sigma_z^{\prime})}
\delta_{\sigma_z^{\prime},\sigma_z^{\prime\prime}}(\delta_{t^{\prime},t^{\prime\prime}}\
Z^{(t^{\prime},t^{\prime\prime})}_0+\delta_{t^{\prime},t^{\prime\prime}+1}\
Z^{(\textbf{k}^\prime,t^{\prime},t^{\prime\prime})}_1)
{B}^{{\textbf{k}^\prime,\textbf{l}}}_{(t^{\prime\prime},\sigma_z^{\prime\prime})}\\\nonumber\\\nonumber&+&\delta_{t,r+1}{B}^{{\textbf{k}^\prime,\textbf{l}}^*}_{(t^\prime,\sigma_z^{\prime})}
\delta_{\sigma_z^{\prime},\sigma_z^{\prime\prime}}(\delta_{t^{\prime},t^{\prime\prime}}\
Z^{(\textbf{k}^\prime,t^{\prime},t^{\prime\prime})}_1
+\delta_{t^{\prime},t^{\prime\prime}+1}\
Z^{(\textbf{k},\textbf{k}^\prime,t^{\prime},t^{\prime\prime})}_3)
{B}^{{\textbf{k}^\prime,\textbf{l}}}_{(t^{\prime\prime},\sigma_z^{\prime\prime})}
],
\end{eqnarray}
where
\begin{eqnarray}
Z^{(t^{\prime},t^{\prime\prime})}_0&\equiv&\frac{2}{N}\sum_{\widetilde{\textbf{R}}^{(t^{\prime},t^{\prime\prime})}}\langle\widetilde{\textbf{R}}^{(t^{\prime},t^{\prime\prime})},\textbf{0}|\widetilde{V}|\textbf{0},\widetilde{\textbf{R}}^{(t^{\prime},t^{\prime\prime})}\rangle,
\\
Z^{(\textbf{k},t^{\prime},t^{\prime\prime})}_1&\equiv&\frac{2}{N}\sum_{i}\sum_{\widetilde{\textbf{R}}^{(t^{\prime},t^{\prime\prime})}}\cos(\textbf{k}\cdot\textbf{p}_i)\
\langle\widetilde{\textbf{p}}_i+\widetilde{\textbf{R}}^{(t^{\prime},t^{\prime\prime})},\textbf{0}|\widetilde{V}|\textbf{0},\widetilde{\textbf{R}}^{(t^{\prime},t^{\prime\prime})}\rangle,
\\
Z^{(\textbf{k},\textbf{k}^\prime,t^{\prime},t^{\prime\prime})}_3&\equiv&\frac{2}{N}\sum_{i,\
j}\sum_{\widetilde{\textbf{R}}^{(t^{\prime},t^{\prime\prime})}}\cos(\textbf{k}\cdot\textbf{p}_i+\textbf{k}^\prime\cdot\textbf{p}_j)\
\langle\widetilde{\textbf{p}}_i+\widetilde{\textbf{R}}^{(t^{\prime},t^{\prime\prime})},\widetilde{\textbf{p}}_j|\widetilde{V}|\textbf{0},\widetilde{\textbf{R}}^{(t^{\prime},t^{\prime\prime})}\rangle.
\end{eqnarray}

 For simplicity, in expression (\ref{e:matrizdirecta})
the Einstein summation convention for the indices
$(t^\prime,t^{\prime\prime},\sigma_z^{\prime},\sigma_z^{\prime\prime})$
is employed. Further, the matrix elements of the exchange potential
are
\begin{eqnarray}\label{e:matrizindirecta}
\\\nonumber G^{ind}_{\textbf{k},(t,\alpha_z),(r,\sigma_z)
}&=&\sum_{\textbf{k}^\prime,\textbf{l}}\Theta_{(\varepsilon_{F}-\varepsilon_{\textbf{l}}(\textbf{k}^\prime))}\times[\
{B}^{{\textbf{k}^\prime,\textbf{l}}^*}_{(r,\sigma_z)}\
S^{(\textbf{k},\textbf{k}^\prime,t,r)}_{0}\
{B}^{{\textbf{k}^\prime,\textbf{l}}}_{(t,\alpha_z)}+{B}^{{\textbf{k}^\prime,\textbf{l}}^*}_{(r,\sigma_z)}\
S^{(\textbf{k},\textbf{k}^\prime,t,r+1)}_{1}\
{B}^{{\textbf{k}^\prime,\textbf{l}}}_{(t+1,\alpha_z)}\\\nonumber\\\nonumber&+&{B}^{{\textbf{k}^\prime,\textbf{l}}^*}_{(r+1,\sigma_z)}\
S^{(\textbf{k},\textbf{k}^\prime,t+1,r)}_{1}\
{B}^{{\textbf{k}^\prime,\textbf{l}}}_{(t,\alpha_z)}+{B}^{{\textbf{k}^\prime,\textbf{l}}^*}_{(r+1,\sigma_z)}\
S^{(\textbf{k},\textbf{k}^\prime,t,r)}_{3}\
{B}^{{\textbf{k}^\prime,\textbf{l}}}_{(t+1,\alpha_z)}],
\end{eqnarray}
where
\begin{eqnarray}
S^{(\textbf{k},\textbf{k}^\prime,t^{\prime},t^{\prime\prime})}_0&=&\frac{2}{N}\sum_{\widetilde{\textbf{R}}^{(t^{\prime},t^{\prime\prime})}}\cos[(\textbf{k}-\textbf{k}^\prime)\cdot\
\textbf{R}^{(t^{\prime},t^{\prime\prime})}]
\times\langle\widetilde{\textbf{R}}^{(t^{\prime},t^{\prime\prime})},\textbf{0}|\widetilde{V}|\textbf{0},\widetilde{\textbf{R}}^{(t^{\prime},t^{\prime\prime})}\rangle,
\\\nonumber&&\\
S^{(\textbf{k},\textbf{k}^\prime,t^{\prime},t^{\prime\prime})}_1&=&\frac{2}{N}\sum_{i}\sum_{\widetilde{\textbf{R}}^{(t^{\prime},t^{\prime\prime})}}\cos[\textbf{k}\cdot\textbf{p}_i+(\textbf{k}-\textbf{k}^\prime)\cdot\
\textbf{R}^{(t^{\prime},t^{\prime\prime})}]
\times\langle\widetilde{\textbf{p}}_i+\widetilde{\textbf{R}}^{(t^{\prime},t^{\prime\prime})},\textbf{0}|\widetilde{V}|\textbf{0},\widetilde{\textbf{R}}^{(t^{\prime},t^{\prime\prime})}\rangle,
\\\nonumber&&\\
S^{(\textbf{k},\textbf{k}^\prime,t^{\prime},t^{\prime\prime})}_3&=&\frac{2}{N}\sum_{i,j}\sum_{\widetilde{\textbf{R}}^{(t^{\prime},t^{\prime\prime})}}\cos[\textbf{k}\cdot(\textbf{p}_i+\textbf{p}_j)+(\textbf{k}-\textbf{k}^\prime)\cdot\
\textbf{R}^{(t^{\prime},t^{\prime\prime})}]
\times\langle\widetilde{\textbf{p}}_i+\widetilde{\textbf{R}}^{(t^{\prime},t^{\prime\prime})},\widetilde{\textbf{p}}_j|\widetilde{V}|\textbf{0},\widetilde{\textbf{R}}^{(t^{\prime},t^{\prime\prime})}\rangle.
\end{eqnarray}
\end{widetext}

\subsection{Reducing the order of some integrals.}
  Any one of the fourth fold  integrals presented  in previous
sections, can be partially integrated  in quadratures, in such a way
they final calculation is reduced to numerically evaluate first
order integrals. Thus
\begin{widetext}
\begin{eqnarray}
\langle\widetilde{\textbf{R}}+\widetilde{\textbf{p}}_1,\widetilde{\textbf{p}}_3|\widetilde{V}|\textbf{0},\widetilde{\textbf{R}}\rangle&=&\frac{\exp[-\frac{1}{2
\widetilde{
a}^2}]}{2\sqrt{2\pi\widetilde{a}^2}}\times\int^{2\pi}_{0}d\phi\
\exp\{-\frac{[(\widetilde{R}_{x_1}+\frac{1}{2})\sin\phi-(\widetilde{R}_{x_2}-\frac{1}{2})\cos\phi]^2}{2\widetilde{a}^2}\}\\\nonumber
&\times&\text{
Erfc}\{-\frac{[(\widetilde{R}_{x_2}-\frac{1}{2})\sin\phi-(\widetilde{R}_{x_1}+\frac{1}{2})\cos\phi]^2}{2\widetilde{a}^2}\},
\end{eqnarray}
where Erfc is the complement error function. Similarly
\begin{eqnarray}
\langle\widetilde{\textbf{R}}+\widetilde{\textbf{p}}_1,\textbf{0}^b|\widetilde{V}|\textbf{0}^b,\widetilde{\textbf{R}}\rangle&=&\frac{\exp[-\frac{1}{4
\widetilde{ a}^2}]}{\sqrt{\frac{4(1+\zeta^2)
\pi\widetilde{a}^2}{\zeta^2}}}\times\int^{2\pi}_{0}d\phi\
\exp\{-\frac{[(\widetilde{R}_{x_1}+\frac{1}{2})\sin\phi-\widetilde{R}_{x_2}\cos\phi]^2}{\frac{(1+\zeta^2)\widetilde{a}^2}{\zeta^2}}
\}\\\nonumber &\times&\text{
Erfc}\{-\frac{[\widetilde{R}_{x_2}\sin\phi-(\widetilde{R}_{x_1}+\frac{1}{2})\cos\phi]^2}{\frac{(1+\zeta^2)\widetilde{a}^2}{\zeta^2}}
\},
\end{eqnarray}
\end{widetext}
where $\zeta\equiv\frac{a}{b}$.

\end{document}